\documentclass[11pt,a4paper]{article}
\usepackage{jcappub}
\usepackage{graphicx}
\usepackage{dcolumn}
\usepackage{bm}
\usepackage{natbib}
\usepackage{afterpage}

\newcommand{\be}{\begin{equation}}
\newcommand{\ee}{\end{equation}}
\newcommand{\bea}{\begin{eqnarray}}
\newcommand{\eea}{\end{eqnarray}}

\newcommand{\s}{{\rm ~s}}

\newcommand{\cm}{{\rm ~cm}}

\newcommand{\kev}{{\rm ~keV}}

\newcommand{\mev}{{\rm ~MeV}}
\newcommand{\GeV}{{\rm ~GeV}}
\newcommand{\gev}{{\rm ~GeV}}
\newcommand{\TeV}{{\rm ~TeV}}
\newcommand{\epm}{{\ensuremath{e^+ e^-\;}}}

\newcommand{\kpc}{{\rm ~kpc}}

\newcommand{\lsim}{\stackrel{<}{\sim}}
\newcommand{\gsim}{\stackrel{>}{\sim}}

\title{Consistent Scenarios for Cosmic-Ray Excesses from Sommerfeld-Enhanced Dark Matter Annihilation}

\author[a,b]{Douglas P. Finkbeiner}
\author[c]{Lisa Goodenough}
\author[a,b,d]{Tracy R. Slatyer}
\author[a]{Mark Vogelsberger}
\author[c,d]{Neal Weiner}

\affiliation[a]{Harvard-Smithsonian Center for Astrophysics, 60 Garden St., Cambridge, MA 02138, USA}
\affiliation[b]{Physics Department, Harvard University, Cambridge, MA 02138, USA}
\affiliation[c]{Center for Cosmology and Particle Physics, Department of Physics, New York University, New York, NY 10003, USA}
\affiliation[d]{School of Natural Sciences, Institute for Advanced Study, Princeton, NJ 08540, USA}

\emailAdd{tslatyer@ias.edu}

\abstract{Anomalies in direct and indirect detection have motivated models of dark matter consisting of a multiplet of nearly-degenerate states, coupled by a new GeV-scale interaction. We perform a careful analysis of the thermal freezeout of dark matter annihilation in such a scenario.  We compute the range of ``boost factors'' arising from Sommerfeld enhancement in the local halo for models which produce the correct relic density, and show the effect of including constraints on the saturated enhancement from the cosmic microwave background (CMB). We find that boost factors from Sommerfeld enhancement of up to $\sim800$ are possible in the local halo.  When the CMB bounds on the saturated enhancement are applied, the maximal boost factor is reduced to $\sim400$ for 1-2 TeV dark matter and sub-GeV force carriers, but remains large enough to explain the observed Fermi and PAMELA electronic signals.  We describe regions in the DM mass-boost factor plane where the cosmic ray data is well fit for a range of final states, and show that Sommerfeld enhancement \emph{alone} is enough to provide the large annihilation cross sections required to fit the data, although for light mediator masses ($m_\phi \lesssim 200$ MeV) there is tension with the CMB constraints in the absence of astrophysical boost factors from substructure.  Additionally, we consider the circumstances under which WIMPonium formation is relevant and find for heavy WIMPs ($\gsim 2 \TeV$) and soft-spectrum annihilation channels it can be an important consideration; we find regions with $m_\chi \gtrsim 2.8 \TeV$ that are consistent with the CMB bounds with $\mathcal{O}$(600-700) present-day boost factors.}


\arxivnumber{1011.3082}

\begin{document}

\maketitle
\flushbottom

\section{Introduction}

Recent measurements of electron and positron cosmic rays at 10-1000 GeV \cite{Barwick:1997ig, AMS, Adriani:2008zr, aticlatest, Torii:2008xu, Abdo:2009zk, Aharonian:2009ah} indicate a rise in the positron flux fraction at 10-100+ GeV and a hardening of the $e^+ + e^-$ spectrum at $\sim 20-1000$ GeV, suggesting a new source of positrons and electrons with a hard spectrum and a TeV-scale cutoff. Annihilation of weak-scale dark matter provides a natural explanation for an excess of pairs at this energy scale, but  conventional models of thermal relic dark matter annihilation are challenged by the large amplitude and hard spectrum of the signal \cite{cirelli,Cholis:2008hb}, as well as the absence of any corresponding excess in cosmic ray antiprotons \cite{Cholis:2008hb, Cirelli:2008pk}.

The presence of a new GeV-scale force in the dark sector has been proposed to explain these features of the measured cosmic-ray excesses \cite{ArkaniHamed:2008qn, Pospelov:2008jd,Nomura:2008ru}. The dark matter can efficiently annihilate to the force carriers $\phi$, which then decay into pairs of Standard Model particles; if $m_\phi < 2 m_p$, the decay into antiprotons is kinematically forbidden, and the resulting $e^+$ spectrum is hard due to the highly boosted intermediate (on-shell) light state \cite{Finkbeiner:2007kk, Cholis:2008vb}. Additionally, $\phi$ mediates a fm-range attractive DM self-interaction, which enhances annihilation at low velocities, potentially by several orders of magnitude, i.e., the Sommerfeld enhancement \cite{sommerfeld}\footnote{This effect was originally studied in the context of dark matter annihilation in \cite{Hisano:2003ec, Hisano:2004ds}.}. The presence of $\mathcal{O}$(MeV) splittings between dark matter states can be used to explain the 511 keV excess in the Galactic center observed by INTEGRAL/SPI \cite{Weidenspointner:2006nu, Weidenspointner:2007}, via collisional excitation and decay of the excited state (``eXciting dark matter'' (XDM)) \cite{Finkbeiner:2007kk, Pospelov:2007xh, Chen:2009dm, Chen:2009av, Finkbeiner:2009mi}, and to reconcile the annual modulation observed by DAMA/LIBRA \cite{Bernabei:1998td,Bernabei:2008yi, Bernabei:2010mq} with the null results of other direct detection experiments through inelastic WIMP-nuclear scattering (``inelastic dark matter'' (iDM)) \cite{Smith:2001hy, Chang:2008gd}.

While the low-velocity enhancement to annihilation has been well explored, and fits to the existing data from the new annihilation channels  have been given \cite{Cholis:2008wq,Meade:2009iu}, no complete picture has yet been presented. The Sommerfeld enhancement is strongly velocity-dependent (scaling as $1/v$ generically), and thus has a much larger effect in the local halo than during freezeout, when the dark matter thermal relic density is established. However, the effect on freezeout can still be significant: an $\mathcal{O}(1)$ enhancement to the annihilation rate at freezeout leads to $\mathcal{O}(1)$ changes in the thermal relic density, so the underlying ``bare'' annihilation cross section must be reduced to compensate \cite{Dent:2009bv, Zavala:2009mi}. In narrow regions of parameter space, where the enhancement scales as $1/v^2$ rather than $1/v$ below a certain velocity, this effect can be more pronounced: in particular, in such ``resonance regions'' it may be possible for DM annihilation to ``un-freeze'' after kinetic decoupling (once the DM begins to cool faster than the CMB), leading to a very efficient depletion of the relic density \cite{ArkaniHamed:2008qn, Feng:2010zp}. Indeed, in cases where $\chi \chi \rightarrow \phi \phi$, $\phi \rightarrow \mu^+\mu^-$, there is significant tension between these bounds and the annihilation rates required to generate the cosmic-ray anomalies \cite{Feng:2009hw, Feng:2010zp}. 

In this work, we study a broader range of parameter space, including a range of decay modes for the $\phi$ and splittings between dark matter states. We will see that while the scenarios discussed by \cite{Feng:2010zp} are tightly constrained, generic models have less tension due to enhanced annihilation in the presence of mass splittings and the typical hard electron component arising from $\phi\rightarrow e^+e^-$. We will employ the approximations derived in \cite{Slatyer:2009vg} to identify regions of interest and present spectra for self-consistent benchmark points that fit the cosmic-ray data; we verify that a careful numerical calculation of the Sommerfeld enhancement at these benchmark points does not appreciably alter our results. We find that the most stringent constraints on the present-day annihilation cross section originate not from achieving the appropriate relic abundance, but from WMAP measurements of the CMB temperature and polarization angular power spectra \cite{Galli:2009zc, Slatyer:2009yq}, which will be greatly improved by Planck. We motivate and describe the models we consider in \S \ref{sec:models}, outline the details of our calculation in \S \ref{sec:history}, and discuss constraints from indirect detection in \S \ref{sec:cmbbounds}. In \S \ref{sec:paramspace} we describe our modeling of the cosmic ray spectra, and our criterion for a ``good fit'' to the cosmic ray data. Readers principally interested in our results can immediately skip to \S \ref{sec:results}, which is essentially self-contained. 

Throughout this work we will use the term ``boost factor'' (BF) to mean the $s$-wave annihilation cross section $\langle \sigma v \rangle$ divided by $3 \times 10^{-26}$ cm$^3$/s (this value has been employed in the literature as the canonical value of $\langle \sigma v \rangle$ for thermal relic DM).

\section{Inelastic Dark Matter Through the Vector Portal}
\label{sec:models}
Let us begin by laying out the model space we consider. We focus here on models where dark matter freezes out by annihilating into a new force carrier $\chi \chi \rightarrow  \phi \phi$ \cite{Finkbeiner:2007kk,Pospelov:2007mp}, which itself maintains kinetic equilibrium with the standard model, typically through the process $e \gamma \leftrightarrow e \phi$ \cite{Finkbeiner:2008gw}, or $s$-channel dark boson exchange \cite{ArkaniHamed:2008qp}. While models can be constructed through a variety of ``portals'', we focus our attention on the vector portal. That is, we assume $\phi$ is the vector boson of a new gauge group $U(1)_D$, Higgsed at roughly the GeV scale, and the connection between dark forces and the standard model comes through an effective term $\epsilon F^\mathrm{EM}_{\mu \nu} F_D^{\mu\nu}$. The dark sector is generically thermalized before dark matter freezeout for $\epsilon \gsim 10^{-7}$, although it should be noted that so long as the sectors are brought into thermal contact through some other physics earlier, this condition could be relaxed \cite{Pospelov:2007mp}.

It is important to emphasize that because we are considering a vector force, we are {\em forced} to consider sectors with multiple states. The minimal fermionic representation of $U(1)_{D}$ is a Dirac fermion, which is composed of two Majorana fermions, while the minimal scalar representation is complex, composed of two real scalars. Once $U(1)_D$ is broken, absent an accidental low-energy global symmetry, there is no a priori reason these states should remain degenerate. If the components are degenerate, then direct detection experiments \cite{Ahmed:2009zw,Aprile:2010um} constrain $\epsilon \lsim 10^{-6}\, (10^{-8})$ for $m_\phi = 1 \gev\, (100 \mev)$. In supersymmetric models, $m_\phi^2 \propto \epsilon$, yielding a fairly robust cross section of $\sim 10^{-38} {\rm cm^2}$ \cite{Cheung:2009qd}, well above current direct detection limits.  However, if the different WIMP components are non-degenerate by an amount $\delta \gsim m_\chi v^2$, then because the vector coupling is off-diagonal \cite{TuckerSmith:2004jv}, the elastic scattering cross section will be suppressed or eliminated.

Thus we should emphasize that within this overall framework {\em the most generic setup is one where dark matter consists of multiple states (i.e., a pseudo-Dirac fermion), split by an amount $\gsim 100 \kev$, and where the force carrier interacts with SM matter through its mixing with the photon.} This final point is especially important, because as we shall see, the splitting naturally enhances the late-time Sommerfeld enhancement, and the coupling of $\phi$ to charge generally produces a sizeable $\phi \rightarrow e^+e^-$ component, which is the dominant contributor to the DM explanation of signals seen at PAMELA and \emph{Fermi}. Other models can be constructed, for instance through the Higgs portal or through the axion portal \cite{Nomura:2008ru}, but degenerate WIMPs annihilating dominantly into $4\mu$ is actually extremely challenging in the context of vector portal models, and not generically realized within this setup.

\subsection{Annihilation channels and Sommerfeld enhancement}
As our canonical example, we consider a model where the dark matter is a Dirac fermion charged under a hidden sector $U(1)_D$, broken at the GeV scale by an Abelian Higgs. At low energies, we assume higher dimension operators will split the Majorana components $\chi_{1,2}$ of the WIMP by an amount $m_{\chi_2}-m_{\chi_1}=\delta \sim \mev$. Even at high energies, it is convenient to think of annihilation processes in terms of these states.

Most generically, the dominant annihilation is $\chi_1 \chi_1,\, \chi_2 \chi_2  \rightarrow \phi \phi$, via $t$-channel exchange of the other $\chi$ state. There are also coannihilation channels into Higgses (which, at late times can be cast in the unitary gauge as $\chi_1 \chi_2 \rightarrow \phi h_D$), or other light states charged under $U(1)_D$.
Both of these channels experience Sommerfeld enhancement. The unenhanced tree-level annihilation cross section into two gauge bosons, for nonrelativistic or mildly relativistic fermions, is given by the usual result for pair production (e.g. \cite{Peskin:1995ev}),
with the dominant low-velocity contribution coming from the $s$-wave term, $\sigma |v_\mathrm{rel}|_{l=0} = \pi \alpha_D^2 / m_\chi^2$. Here $v_\mathrm{rel}$ is the relative velocity of the two interacting particles, and $\vec{p}$ is the 3-momentum of a single initial fermion in the COM frame.

The cross section for $s$-channel coannihilation into two charge-1 dark Higgses is given by,
\begin{equation} \sigma = \frac{\pi \alpha_D^2}{8} \frac{m_\chi^2 + \frac{2}{3} |\vec{p}|^2}{|\vec{p}| \left(m_\chi^2 + |\vec{p}|^2 \right)^{3/2}},\end{equation}
and scales as the Higgs charge squared. The leading $s$-wave contribution is $\sigma |v_\mathrm{rel}|_{l=0} = \pi \alpha_D^2 / 4 m_\chi^2$, a factor of 4 lower than for the $t$-channel annihilation (for a singly charged Higgs).

To calculate the Sommerfeld enhancement for these channels, one must sum a series of ladder diagrams (or equivalently, solve the Schr\"odinger equation).
The interaction between the dark gauge bosons and the $\chi_i$ eigenstates is off-diagonal, so an initial two-particle state $\chi_1 \chi_1$ can be scattered into the two-particle state $\chi_2 \chi_2$ (or vice versa), but not into the state $\chi_1 \chi_2$. Similarly, for particles initially in a two-body $\chi_1 \chi_2$ state, the only effect of the long-range vector interaction is to swap the individual particle states. Thus, when computing the potential due to the long-range interaction, the $\chi_1 \chi_2$ 2-body state is disjoint from the other two and can be treated separately.

Since the vector-mediated scatterings are always purely elastic for the $\chi_1 \chi_2$ 2-body state, and the long-range interaction is attractive,
the effect of the Sommerfeld enhancement is well approximated by the enhancement due to a Yukawa potential (up to corrections of order $\delta / m_\chi$, due to the slightly different masses of the interacting particles).

For the $\chi_1 \chi_1$ and $\chi_2 \chi_2$ states, the interaction at tree-level is always inelastic. The corresponding scattering problem can be written in terms of a potential matrix with off-diagonal Yukawa terms, and approximately solved as in \cite{Slatyer:2009vg}. We employ the approximate semi-analytic results of \cite{Slatyer:2009vg} to estimate the Sommerfeld enhancement; for our benchmark points, we also check the approximation numerically.

\subsection{A specific model for the mass splitting}
\label{sec:masssplittingann}

In the case where the Higgs is singly charged, an $\mathcal{O}$(MeV) mass splitting can be generated naturally by a higher-dimension operator, which in the high-energy theory (where the $U(1)_D$ is unbroken) takes the form $(1/2)(y/\Lambda) (\bar{\Psi^C} \Psi h_D^* h_D^* + h.c)$. In Appendix \ref{sec:makesplitting} we calculate the mass splitting and annihilation rate arising from this operator; here we will summarize those results.

In terms of the $\chi_{1,2}$ mass eigenstates, the operator takes the form,
\[ \mathcal{L}_\mathrm{split} = (1/4) (y/\Lambda) (\chi_1 \chi_1 (h_D h_D + h_D^* h_D^*) - \chi_2 \chi_2 (h_D h_D + h_D^* h_D^*) + 2 i \chi_1 \chi_2 (h_D h_D - h_D^* h_D^*)) + h.c. \]
Working in unitarity gauge for simplicity, and writing $h_D \rightarrow (v_D + h_D)/\sqrt{2}$, we obtain,
\[ \mathcal{L}_\mathrm{split} \rightarrow (1/4) (y/\Lambda) (\chi_1 \chi_1 - \chi_2 \chi_2) (v_D^2 + h_D^2 + 2 v_D h_D) + h.c. \] 
The Yukawa coupling is suppressed by the small ratio $v_D / \Lambda$, but the $(1/4) (y/\Lambda) \chi_i \chi_i h_D^2$ term induces a potentially large $\chi_i \chi_i$ annihilation channel, as discussed in \cite{Ruderman:2009tj}. Furthermore, the mass splitting $\delta = (y/\Lambda) v_D^2$, and the mediator mass $m_\phi = g_D v_D$: this allows us to write the low-velocity cross section for $\chi_i \chi_i \rightarrow h_D h_D$ annihilation as,
\begin{equation} \sigma v_\mathrm{rel} = \frac{1}{2} v^2 \left( \frac{\delta m_\chi }{m_\phi^2}\right)^2 \frac{\pi \alpha_D^2}{m_\chi^2}. \end{equation}

This relation comes from a tree-level computation and neglects Sommerfeld corrections. Since the $\chi_i \chi_i \rightarrow h_D h_D$ channel corresponds to annihilation of two same-charge fermions in the high-energy limit, it is actually Sommerfeld \emph{suppressed}: in the ladder diagram picture, this can be understood as due to destructive interference between the two relevant diagrams (with the annihilation being through $\chi_1 \chi_1$ or $\chi_2 \chi_2$ in the final step of the ladder diagram), originating from the fact that the $\chi_1 \chi_1 h_D h_D$ and $\chi_2 \chi_2 h_D h_D$ couplings have opposite signs. Since Sommerfeld-suppressed channels are negligible for low $v$, we will approximate the Sommerfeld suppression by the result for $p$-wave Coulomb scattering, 
\begin{equation} S_\mathrm{rep,p-wave} = \left(\frac{\pi / \epsilon_v}{e^{\pi/\epsilon_v} - 1} \right) \left(1 + \frac{1}{4 \epsilon_v^2} \right). \end{equation}

This channel is therefore always negligible in the present day, as it experiences both a Sommerfeld suppression and a $p$-wave suppression. However, it can be important at freezeout, depending on the ratio $m_\chi \delta / m_\phi^2$. If the DM were instead a complex scalar, the $p$-wave suppression would be absent and the effect on freezeout much larger\footnote{We thank Josh Ruderman for this observation.}.

\section{Solving for the Relic Density}
\label{sec:history}

We solve numerically for the abundances in the ground (1) and excited (2) states, using the publicly available \texttt{IDL} code \texttt{LSODE}. We define $Y_i = n_i/s$, where $s$ is the entropy density, and $x = m_\chi / T$. The Boltzmann equation becomes,
\begin{align} \frac{d Y_1}{d x} & = \frac{x}{H(m_\chi)} \left(\Gamma Y_2 -s \left[ (Y_1^2 - Y_\mathrm{eq}^2) \langle \sigma^A_{11} v \rangle + (Y_1 Y_2 - Y_\mathrm{eq}^2) \langle \sigma^A_{12} v \rangle + k_E Y_1^2 - k_D Y_2^2 \right] \right), \nonumber \\
\frac{d Y_2}{d x} & =  \frac{x}{H(m_\chi)} \left(- \Gamma Y_2 -s \left[ (Y_2^2 - Y_\mathrm{eq}^2) \langle \sigma^A_{22} v \rangle + (Y_1 Y_2 - Y_\mathrm{eq}^2) \langle \sigma^A_{12} v \rangle + k_D Y_2^2 - k_E Y_1^2 \right]  \right). \end{align}
Here $\Gamma$ is the decay rate for the excited state $\chi_2$, and $k_E$ and $k_D$ describe (respectively) the excitation and de-excitation of the excited state by DM-DM scattering. We include the $s$- and $p$-wave contributions to the annihilation cross sections. We have verified that including the complete (tree-level) relativistic cross sections, with the $s$- and $p$-wave pieces corrected by Sommerfeld enhancement, and using the full relativistic momentum distribution for thermal particles (rather than the nonrelativistic Maxwell-Boltzmann distribution), does not significantly alter our results.

$Y_\mathrm{eq}$, the equilibrium value of the abundances (normalized to the entropy density), is determined simply by the number of degrees of freedom of the DM compared to the SM relativistic degrees of freedom, and the temperature and mass of the DM. Strictly the ground and excited states have distinct values of $Y_\mathrm{eq}$, but due to the large hierarchy between the dark symmetry breaking scale and the dark matter mass, the splitting between the states is irrelevant during freezeout. By the time the temperature of the universe drops to the symmetry breaking scale, much less the scale of the mass splittings, $Y_\mathrm{eq}$ is infinitesimal and irrelevant to the evolution of $Y_1$, $Y_2$. Thus it is acceptable to approximate $Y^1_\mathrm{eq} = Y^2_\mathrm{eq}$. 

The Weyl fermions $\chi_1$ and $\chi_2$ each have $g=2$ internal degrees of freedom, and we use the standard results \cite{kolbandturner},
\begin{equation} Y_\mathrm{eq} = \left\{  \begin{array}{cc} \frac{45}{2\pi^4} \left(\frac{\pi}{8} \right)^{1/2} \frac{g}{g_{*S}} x^{3/2} e^{-x}, & x \gg 3, \\ \frac{45 \zeta(3)}{2 \pi^4} \frac{g_\mathrm{eff}}{g_{*S}}, & x \ll 3. \end{array} \right. \label{eq:yeq} \end{equation}
Here $g_\mathrm{eff} = 3g/4$ for the fermionic dark matter we consider; for bosonic DM $g_\mathrm{eff} = g$. $g_{*S}$ counts the relativistic degrees of freedom for the entropy density. 

Once the DM thermally decouples from the SM photon bath, it cools more rapidly and the Sommerfeld enhancement becomes more pronounced \cite{ArkaniHamed:2008qn, Dent:2009bv, Zavala:2009mi}. Prior to kinetic decoupling, the DM temperature is equal to the CMB temperature, scaling as $g_{*S}^{-1/3}/a$; after decoupling, it scales as $1/a^2$. Thus we obtain the relation, 
\[T_\mathrm{DM} = \frac{T_\mathrm{CMB}^2}{T_{KD}} \left(\frac{g_{*S}(T_\mathrm{CMB})}{g_{*S}(T_{KD})} \right)^{2/3}. \]

We employ the expression for the kinetic decoupling temperature derived in \cite{Feng:2010zp}, with the added requirement that the kinetic decoupling temperature must be larger than the mass splitting, so that inelastic scatterings of DM on SM particles are not kinematically suppressed,
\begin{equation} T_\mathrm{kd}^e \sim \mathrm{max} \left\{m_e, \delta, 0.82 \mathrm{MeV} \left[\frac{10^{-3}}{\epsilon} \right]^{1/2} \times 
\left[\frac{m_\phi}{30 \mathrm{MeV}} \right]  \left[\frac{0.021}{\alpha_D} \right]^{1/4} \left[\frac{m_\chi}{\mathrm{TeV}} \right]^{1/4} \right\}. \end{equation}
We take $\epsilon = 10^{-3}$ as a benchmark, but our results are not very sensitive to this choice.

We have neglected the terms in the Boltzmann equation describing the up- and down-scattering of DM states on SM particles: at temperatures where the mass splitting becomes significant and the ground and excited states may have different populations, these processes are (1) slow compared to a Hubble time, and (2) subdominant compared to the effects of DM-DM interactions, in models with Sommerfeld-enhanced scattering. (1) follows simply from the observation that kinetic decoupling occurs at temperatures higher than the mass splitting; (2) can be seen e.g. from \cite{Feng:2010zp}, where DM-DM interactions are found to become inefficient at preserving thermal equilibrium at $\sim 10$ keV, well below the kinetic decoupling temperature.

The excitation fraction affects the annihilation rate in two ways. The smaller effect is that the Sommerfeld enhancements differ for $\chi_1 \chi_1$, $\chi_2 \chi_2$ and $\chi_1 \chi_2$ initial states (although of course they are identical in the limit where the temperature greatly exceeds the mass splitting).

More importantly, the annihilation cross section for a $\chi_1 \chi_2$ initial state differs from that for a $\chi_1 \chi_1$ or $\chi_2 \chi_2$ initial state, so the annihilation cross section averaged over all possible pairs of interacting DM particles is a function of the excitation fraction. Let the total DM number density be denoted $n$, and the number densities for the ground and excited states be denoted $n_1$ and $n_2$ respectively. Then the number of annihilations per volume per second, in the absence of Sommerfeld enhancement, is given for this simple model by $n_1^2 \langle \sigma^A_{11} v \rangle / 2 + n_2^2 \langle \sigma^A_{22} v \rangle / 2 + n_1 n_2 \langle \sigma^A_{12} v \rangle$. The first two terms come from $t$-channel annihilation into $\phi$'s, the last term from $s$-channel annihilation into dark Higgses. Let us write $\sigma \equiv \sigma^A_{11} = \sigma^A_{22} = 4 \sigma^A_{12}$, taking the $s$-wave annihilation cross sections derived previously. Then at freezeout, $n_1 \approx n_2 \approx n/2$ and the total annihilation rate per unit volume is given by,
\[ \langle \sigma v \rangle ((n/2)^2 + (1/4)(n/2)^2) = \frac{5}{16} \langle \sigma v \rangle n^2. \]
In the present day, $n_2 \approx 0$ and $n_1 \approx n$, so the annihilation rate per unit volume is $\langle \sigma v \rangle n^2/2$; larger by a factor of $k = 8/5$. More generally, a lower excitation fraction leads to a greater annihilation rate. However, this statement is entirely a function of the ratio of the $\chi_1 \chi_1$, $\chi_1 \chi_2$ and $\chi_2 \chi_2$ annihilation cross sections: in particular, if there are other states present in the dark sector that couple to the dark gauge boson, the $s$-channel $\chi_1 \chi_2$ annihilation is significantly enhanced due to the greater number of final states, leading to a reduction or even reversal of this effect. If the $s$-channel $\chi_1 \chi_2$ annihilation cross section is given by $\langle \sigma^A_{12} v \rangle_{l=0} = \kappa \pi \alpha_D^2 / m_\chi^2$, the rescaling factor $k$ becomes $k = 2/(1+\kappa)$. Note, however, that such additional or enhanced $\chi_1 \chi_2$ annihilation channels also modify the value of $\alpha_D$ required to obtain the correct relic density, and therefore change the Sommerfeld enhancement: the present-day annihilation cross section cannot simply be rescaled by $k$ relative to the case with $\kappa = 1$. We will focus on the minimal $\kappa = 1/4$ case corresponding to a singly charged Higgs, with a $\chi \chi h h$ operator generating the mass splitting (and giving rise to an additional annihilation channel) as described in \S \ref{sec:masssplittingann}, but we will also present results for $\kappa = 1$ and 4 for illustrative purposes. For these latter cases we do not introduce an explicit mechanism to generate the mass splittings; we assume the only contribution to the self-annihilation cross section arises from the usual $\chi \chi \rightarrow \phi \phi$ channel, and parameterize the coannihilation cross section by $\kappa$.

\subsection{DM-DM scattering}
\label{subsec:scattering}

DM-DM scattering does not affect the relic density of dark matter directly, but when the temperature of the universe drops below the mass splitting, the excitation and de-excitation rates can affect the relative populations of the ground and excited states, which in turn impacts the annihilation rate. However, because of the large hierarchy between the DM mass and the mass splittings, these effects only come into play long after freezeout, and thus have little effect on the relic density except very close to resonances. However, we include them for completeness.

We are interested in regions of parameter space where a large Sommerfeld enhancement persists at low velocities; consequently, the mass splitting cannot be much greater than the Rydberg energy of the $\chi \chi$ system \cite{Slatyer:2009vg}. It follows that at low velocities the dominant contribution to the DM-DM scattering arises from infinite ladder diagrams; the principal differences between elastic and inelastic scattering (or excitation vs de-excitation) are then just the transferred momentum ($q^2 \ge m_\chi \delta$ for inelastic scattering) and the final-state phase space (since the ladders differ only in the masses of the initial- and final-state particles, and we assume $\delta \ll m_\chi$). Therefore we expect the elastic scattering, excitation and de-excitation cross sections to be roughly equal, up to phase space factors and with the replacement $m_\phi \rightarrow \sqrt{m_\chi \delta}$ if $m_\chi v \lesssim \sqrt{m_\chi \delta} \lesssim m_\phi$. For the elastic scattering cross section we employ the prescription of \cite{Feng:2010zp} for the momentum transfer cross section, based on \cite{PhysRevLett.90.225002},
\begin{equation}
\sigma_T \approx \left\{ \begin{array}{cc} \frac{4\pi}{m^2_\phi} \beta^2 \ln(1+\beta^{-1}),
& \beta<0.1, \\
\frac{8\pi}{m^2_\phi}
\beta^2/(1+1.5\beta^{1.65}),
& 0.1 \le \beta \le 1000, \\
\frac{\pi}{m^2_\phi}
\left( \ln\beta + 1 - \frac{1}{2} \ln^{-1} \beta \right)^2,
& \beta > 1000,
\end{array} \right.
\end{equation}
where $\beta = 2 \alpha_D m_\phi / (m_\chi v_\mathrm{rel}^2)$.

\subsection{Decay of the excited state}

In our current toy model, for mass splittings much smaller than $2 m_e$, the lifetime of the excited state exceeds the age of the universe \cite{Finkbeiner:2009mi}. This is potentially problematic due to constraints from direct detection on the present-day excitation fraction (since the excited state can downscatter in detectors) \cite{Finkbeiner:2009mi}: consequently, for $\delta \ll 2 m_e$, either the excited state must be efficiently depleted by DM-DM scattering, or the model must have some extra ingredient that can mediate the decay of the excited state. A lifetime between $10^{13}-10^{18}$ s may be of particular interest, as that would allow the annihilation rate at the redshift of last scattering to be reduced relative to the present day, since the DM annihilates more efficiently when it is entirely in the ground state (at least for models with $\kappa < 1$), weakening the CMB constraints. Once the mass splitting exceeds $2 m_e$, opening up decay to an $e^+ e^-$ pair, the lifetime of the excited state is naturally around that of the neutron, or $\sim 10^3$ s for $\epsilon \sim 10^{-3}$ (the lifetime scales as $1/\epsilon^2$) \cite{Finkbeiner:2009mi}. In any case, these lifetimes greatly exceed the age of the universe at kinetic decoupling, and \emph{cannot significantly affect the relic density}. For the purposes of this calculation, we set $\Gamma = 6.6 \times 10^{-28}$ GeV, corresponding to a lifetime of $10^3$ s, as appropriate for a mass splitting greater than $2 m_e$ with $\epsilon \sim 10^{-3}$. We have verified that our results are insensitive to the decay lifetime, at least for the relatively short lifetimes relevant for $\delta \gtrsim 2 m_e$: increasing the lifetime up to $10^9$ s (corresponding to $\epsilon \sim 10^{-6}$) has no measurable impact on our results, and increasing the lifetime to infinity affects our results only in the sense that annihilation in the present day can also involve excited-state DM (that is, it affects the present-day annihilation rates relevant for indirect detection, but does not modify the relic density).

\section{Constraints from the Cosmic Microwave Background and Other Indirect Searches}
\label{sec:cmbbounds}

Many bounds on dark matter annihilation from indirect detection searches have been discussed in the literature. However, most of them have large uncertainties associated with model-dependent or poorly known astrophysical factors, and so we do not employ them as constraints on our allowed regions of parameter space; later in this section we will briefly outline these bounds and justify neglecting them. The exception is a set of constraints from the cosmic microwave background that provide uniquely clean bounds on the dark matter annihilation cross section in models with Sommerfeld-enhanced annihilation, with no dependence on Galactic astrophysics or the history of DM structure formation, and minimal dependence on the details of the DM model; we now give a brief outline of these constraints and their application.

Dark matter annihilation around the redshift of last scattering can significantly perturb the temperature and polarization angular power spectra of the cosmic microwave background (CMB) \cite{Padmanabhan:2005es, Galli:2009zc, Slatyer:2009yq}\footnote{Limits from the CMB spectrum itself are much weaker than the temperature and polarization anisotropy constraints \cite{Zavala:2009mi, Hannestad:2010zt}.}. DM models with Sommerfeld-enhanced annihilation are especially sensitive to the resulting constraints, since the annihilation cross section increases at low velocity. The $95\%$ confidence limits on DM annihilation from WMAP5 can be expressed as \cite{Slatyer:2009yq}\footnote{Note that these constraints assume a constant primordial scalar spectral index $n_s$ in re-fitting the cosmological parameters; if running of the spectral index is allowed, these constraints may be weakened somewhat.},
\begin{equation} \frac{\lim_{v \rightarrow 0} \langle \sigma v \rangle}{3 \times 10^{-26} \mathrm{cm}^3/\mathrm{s}} \lesssim \frac{120}{f} \left( \frac{m_\chi}{1 \mathrm{TeV}} \right).\end{equation} 
Here $f$ is a parameter describing the fraction of energy from DM annihilation which ionizes and heats the intergalactic medium;  \cite{Slatyer:2009yq} showed that $f$ can be approximated as $f \sim 0.7$ for annihilation to electrons and $f \sim 0.2-0.3$ for all other SM final states, excepting neutrinos, where $f \sim 0$ (that is, $f \sim 0$ if the DM annihilates directly to neutrinos; if the DM annihilates to unstable SM particles which then decay to neutrinos, the value $f \sim 0.2-0.3$ should be used). Thus provided DM annihilations have no significant branching ratio directly to neutrinos, the boost factor at very low velocity is constrained to satisfy BF$(v \rightarrow 0) \lesssim 600 m_\chi / 1$ TeV, with a smaller limit if the dark sector annihilation products decay to electrons with a non-negligible branching ratio.

This constraint can be combined with the semi-analytic expression for the Sommerfeld enhancement to obtain a DM-mass-independent limit on the allowed boost factor in the present day. Using the semi-analytic form for the Sommerfeld enhancement derived in \cite{Slatyer:2009vg}, as $v \rightarrow 0$ the enhancement saturates at,
\begin{equation} S = \left(\frac{2 \pi^2 }{\mu}\right) \frac{1}{1-\cos\left(\epsilon_\delta \pi /\mu+2 \theta_-\right)} > \frac{\pi^2 }{\mu}. \end{equation}
Here $\mu$ is a function of $\epsilon_\phi \equiv m_\phi / \alpha_D m_\chi$ and $\epsilon_\delta \equiv \sqrt{2 \delta / m_\chi} / \alpha_D$ (defined in \cite{Slatyer:2009vg}), but in practice it is nearly independent of $\epsilon_\delta$: for the region where the approximations of \cite{Slatyer:2009vg} are expected to be accurate ($\delta \ll \alpha_D m_\phi \ll \alpha_D^2 m_\chi$), we find $\mu \approx (1/2) (1 + \sqrt{5}) m_\phi / \alpha_D m_\chi$ to within $30\%$.

In the present day, if $\mu \ll \epsilon_v$, then the enhancement is bounded above by $S \lesssim 2 \pi / \epsilon_v$ (in the inelastic case, $S \approx 2 \pi / \epsilon_v$, whereas in the elastic case $S \approx \pi / \epsilon_v$). Then the ratio of the present-day enhancement to the saturated enhancement is bounded by,
\begin{equation} \frac{S_\mathrm{now}}{S_\mathrm{sat}} \lesssim \frac{2 \pi / \epsilon_v}{\pi^2 / \mu} = \frac{2}{\pi} \left( \frac{\mu}{\epsilon_v}\right) = \frac{2}{\pi} \left(\frac{ m_\mu}{ m_\chi v}\right), \end{equation}
where $m_\mu = \alpha_D m_\chi \mu$. The limit from WMAP5 requires that $BF_\mathrm{sat} < (120/f) m_\chi / 1 \mathrm{TeV}$, which in turn yields a limit on the present-day boost factor depending only on $m_\phi, \delta$ and the local velocity dispersion:
\begin{equation} BF_\mathrm{now} \lesssim (2/\pi)  (120/f) (m_\mu/ 1 \mathrm{GeV}) (10^{-3}/v). \end{equation}
Taking $v \sim 5 \times 10^{-4}$, and writing $m_\mu = (1/2) (1 + \sqrt{5}) m_0$ (so that $m_0 \sim m_\phi$, but with some additional dependence on $\delta$ and $m_\phi$), we obtain,
\begin{equation} BF_\mathrm{now} \lesssim (250/f) (m_0/ 1 \mathrm{GeV}). \end{equation}
For example, taking $f \sim 0.6-0.7$, as appropriate for the pure-electron case, yields a maximal present-day boost factor of 70-80, for $m_0 \sim 200$ MeV.

\subsection{Other limits from the early universe}

Limits on DM annihilation from the total optical depth to the last scattering surface were studied by \cite{Huetsi:2009ex, Cirelli:2009bb}\footnote{The possibility for DM to contribute significantly to reionization had been previously discussed in \cite{Belikov:2009qx}.}. These constraints are extremely similar in spirit to the CMB limits studied above, measuring the perturbation to the optical depth from the modified ionization history, and would seem to be subsumed in the complete likelihood analysis taking into account the modified ionization history (which gives rise to the limits quoted above); the comparison of the integrated optical depth (due to a nonstandard ionization history) directly to the WMAP limit on $\tau$ (which is derived assuming the standard ionization history) may introduce some error relative to the complete likelihood analysis. The limits in \cite{Cirelli:2009bb} appear stronger by a factor of $\sim 2$: however, their approach tends to overestimate (by an $\mathcal{O}(2)$ factor) the fraction of deposited energy relative to the careful numerical calculation \cite{Slatyer:2009yq}. Limits from the extra heating induced by DM annihilation in the early universe were also studied by \cite{Cirelli:2009bb}: they were found to be generically weaker than the optical depth bounds.

Another possible early-universe limit can be obtained from the extragalactic gamma-ray background \cite{Profumo:2009uf,Belikov:2009cx,Huetsi:2009ex,Abazajian:2010sq,Hutsi:2010ai}, but this constraint tends to depend very strongly on the assumed DM structure formation history, and the density profiles of the early halos. Depending on the halo density profile and the extrapolation of the halo mass function to low masses, this bound can be an order of magnitude \emph{either} stronger or weaker than the limits derived from the CMB; thus, it does not provide any additional robust constraints on the scenarios we consider. Furthermore, the DM self-interactions present in models of this type, and velocity ``kicks'' from decays or de-excitations of the excited state, both have the potential to modify structure formation in ways which are not accounted for in collisionless cold dark matter simulations (see e.g. \cite{Loeb:2010gj, Bell:2010qt} and references therein). This issue also affects potential constraints from DM substructure in the Milky Way.  

\subsection{Limits from the Galactic halo}

Measurements of gamma-rays and neutrinos from the Galactic halo have also been used to constrain DM annihilation. The gamma-ray signal has two components: (1) photons produced directly in DM annihilation by final state radiation, or by production of neutral pions which decay into gammas, and (2) the signal produced by inverse Compton scattering of high-energy electrons on starlight, infrared and CMB photons. Signals from neutrinos and the first gamma-ray component (which we will denote ``FSR'') depend on the dark matter density profile and (because of Sommerfeld enhancement) the dark matter velocity profile; the second gamma-ray component (denoted ``ICS'') has additional astrophysical uncertainties associated with the electron propagation.

The neutrino signal, while somewhat model-dependent (requiring a significant branching ratio into non-$e^+e^-$ final states), has the advantage of having no significant astrophysical background; future studies at IceCube may place very robust constraints on the scenarios we consider \cite{Spolyar:2009kx,Mandal:2009yk}. However, for e.g. a pure 4$\mu$ final state, the present limits from SuperKamiokande can only rule out higher-mass DM ($m_\chi \gtrsim 3$ TeV) models fitting the cosmic-ray excesses, assuming an Einasto DM density profile and no significant change in the DM velocity distribution with Galactocentric radius \cite{Meade:2009iu}; even accepting these (far from conservative) assumptions, most of the parameter space we will be interested in lies at lower DM masses.

The gamma-ray bounds \cite{Bell:2008vx,Bertone:2008xr,Bergstrom:2008ag,Cirelli:2009vg,Pato:2009fn,Meade:2009iu,Cirelli:2009dv,Papucci:2009gd,Hutsi:2010ai} are often more constraining\footnote{It is worth noting, however, that bounds from the H.E.S.S measurement of gamma-ray emission from the Galactic Ridge rely on a background subtraction which can also include DM-related emission, and this has not always been taken into account in setting limits \cite{Mack:2008wu}.}. However, the most recent conservative analyses \cite{Cirelli:2009dv,Papucci:2009gd,Hutsi:2010ai} indicate that DM annihilation models fitting the cosmic-ray data remain allowed if the final state consists of electrons and muons, and the DM density profile is rather shallow; these models are generally not ruled out by any Galactic gamma-ray limits (models with a mixture of electrons, muons and charged pions in the final state were not tested in these papers, but since the spectrum is fairly similar we do not expect the constraints to change greatly). Shallow DM profiles, like the ``cored isothermal'' or Burkert benchmarks commonly used in the literature, are disfavored by DM-only N-body simulations, but appear to agree better with observations (see e.g. \cite{2010AdAst2010E...5D} for a recent review). It is not clear how the inclusion of baryons affects the DM density profile in the inner galaxy: it has been suggested that baryons could either steepen the DM cusp via adiabatic contraction \cite{1986ApJ...301...27B,2004ApJ...616...16G} or flatten it via dynamical friction \cite{ElZant:2001re,ElZant:2003rp}.

In Sommerfeld-enhanced models, the uncertainty in the density profile is compounded by uncertainty in the variation of the velocity distribution with radius. If the DM velocity drops markedly in the Galactic center, as generically expected from DM-only simulations (e.g. \cite{Navarro:2008kc}), the constraints can become more stringent, as the Sommerfeld enhancement becomes larger (if it is not yet saturated); the shift in gamma-ray constraints in this scenario has been considered by \cite{Cirelli:2010nh}. Conversely, if the velocity dispersion of DM particles rises in the inner halo, the annihilation cross section will decrease, relaxing the constraints \cite{Cholis:2009va}. Some simulations including baryons find a rise in the velocity dispersion in the inner galaxy, accompanied by a flattened DM core \cite{fabioprivate,RomanoDiaz:2008wz,RomanoDiaz:2009yq,Abadi:2009ve,Pedrosa:2009rw,Tissera:2009cm}. 

Increased tidal disruption is expected to lower the amount of substructure closer to the Galactic center (e.g. \cite{Diemand:2006ik,Springel:2008cc,Pieri:2009je}), so if there is a substantial contribution to the PAMELA and \emph{Fermi} signals from DM substructure in the neighborhood of the solar system, this could alleviate tension between scenarios fitting the cosmic-ray data and both the CMB constraints and the gamma-ray limits from the inner Galaxy \cite{Lattanzi:2008qa,Kamionkowski:2010mi,Cline:2010ag,Vincent:2010kv}. Our assumption of a Maxwell-Boltzmann distribution for the local DM velocity is also only an approximation to the reality: instead employing a velocity distribution based on N-body simulations can increase the Sommerfeld-enhanced annihilation rate by up to a factor of $\sim 2$ \cite{privcommLisanti}, which improves the range of parameters consistent with the CMB that can also explain the cosmic-ray data, and may (depending on the variation in the shape of the velocity distribution with Galactocentric radius) also modify the Galactic gamma-ray limits.

Furthermore, in the scenarios we consider -- with their large branching ratios to leptons and small direct photon production -- the strongest limits tend to come from inverse Compton scattering, and astrophysical uncertainties on the cosmic ray propagation are relevant. Most of the analyses to date have assumed that essentially all of the electron energy is lost to ICS, rather than to synchrotron radiation; however, this assumption is strongly dependent on the magnetic field in the inner Galaxy, which is not well constrained (for a recent review of B-fields in the Galactic Center see \cite{Ferriere:2009dh}, and \cite{Sun:2007mx} for a discussion of fields off the Galactic plane). Furthermore, \emph{Fermi} data indicate the presence of a large-scale structure in gamma-rays toward the Galactic center, extending roughly 50$^\circ$ north and south of the plane \cite{Su:2010qj}. It has been suggested that this structure could be associated with an AGN jet or Galactic wind; if this is the case, cosmic ray propagation in this region -- which includes the region used to set some of the most stringent constraints -- could be greatly modified, and ICS limits from this region assuming steady-state diffusive propagation could be very inaccurate (for example, electrons -- both background cosmic rays and DM annihilation products -- may be swept away by the jet or wind before they can scatter).

\subsection{Limits on the force carrier mass}

Measurements of gamma-ray emission from dwarf galaxies have been used to constrain Sommerfeld-enhanced models specifically, since the velocity dispersion in these structures is thought to be lower than in the local halo \cite{Essig:2010em}; at present, mediator masses $m_\phi \lesssim 100$ MeV are disfavored by these constraints. Constraints on the dark matter self-interaction cross section have been discussed recently by \cite{Buckley:2009in,Feng:2009hw}, with the result that mediator masses $m_\phi \lesssim 30-40$ MeV can be excluded. Consequently, we choose all our benchmark mediator masses to exceed 100 MeV.

\section{The Parameter Space for PAMELA and \emph{Fermi}}
\label{sec:paramspace}
With the results of the previous section, we can now answer the question: for what, if any, parameter choices can the cosmic ray signals of PAMELA and \emph{Fermi} be explained, while still achieving the appropriate relic abundance and evading the CMB limits? Determining whether such regions of parameter space exist can be somewhat involved, because of the interplay between the various parameters. Specifically, the particular Sommerfeld enhancement depends greatly on $m_\phi$, but so too does the decay mode of $\phi$, and thus the needed cross section as well. As such, we approach this in a systematic fashion. First, for a representative set of decay modes for $\phi$, spanning a range of masses $m_\phi = 200-900$ MeV, we determine the appropriate ranges of cross section and mass that can explain the data. Having found the preferred values, we then study the Sommerfeld enhancement along slices of fixed $m_\phi$ and $m_\chi$, keeping the relic abundance fixed to the observed value. Finally, we select ``benchmark points'' in this parameter space to give representative examples of explicit fits to the data.

To determine whether a particular choice of $(m_\chi, m_\phi, \mathrm{BF})$ explains the \epm signals, we consider a broad range of parameters describing the background \epm  spectra and cosmic ray propagation.  We use the publicly available \texttt{GALPROP} code \cite{Strong:1999sv} to calculate the local cosmic ray (CR) signals as measured by \emph{Fermi} and PAMELA.  \texttt{GALPROP} allows for considerable freedom in the details of cosmic ray production and propagation.  Among other things, one may choose 
the electron and proton injection spectra, i.e. the spectra of primary electrons and protons emitted at the source;
the diffusion parameters, which include the diffusion coefficient and the size of the cylindrical diffusion zone;
and the Galactic magnetic field and interstellar radiation field, which determine cosmic ray energy losses.

\subsection{Background cosmic ray spectra}
\label{sec:BGspectra}
The background of CR electrons consists of ``primary" and ``secondary" electrons; primary electrons are injected into the Galaxy by cosmic ray sources such as supernova remnants, while secondary electrons are created in the interactions of CR protons with the interstellar gas.  The background of positrons is entirely secondary in nature, arising from the interactions of the CR protons with the interstellar medium.  Therefore, both the primary electron spectrum and the primary proton spectrum are fundamental components of our analysis of the fits to PAMELA and \emph{Fermi}.  Both the proton and electron injection spectra are broken power laws in energy and can be described by four parameters each $(n, \gamma_{1}, \gamma_{2}, E_b)$, an overall normalization $n$, the power law indices below and above the break $\gamma_{1}$ and $\gamma_{2}$, and the energy at which the break occurs $E_b$\footnote{The \texttt{GALPROP} code describes the primary electron spectrum by a double broken power law in energy, which requires two additional parameters, a second break energy and a third power law index.  The spectrum typically has two breaks above $\sim100$ MeV, one around a few GeV and a second around 1 TeV.  The low energy break results from the energy dependence of the dominant energy loss processes changing from $\propto1/E$ to $\propto1/E^2$.  A high energy break is expected to exist due to the finite distance CR electrons can travel from the source before losing a large portion of their energy.  The distances to the nearest CR sources, for example supernova events, determine the highest energy the background electrons reaching our detectors can have.  Since our PAMELA and \emph{Fermi} fits do not extend to energies below 10 GeV, the low energy electron spectrum is irrelevant.  Therefore, for our purposes, the primary electron spectrum can be described by a broken power law with two indices and one break energy.}.

We constrain the primary proton spectrum by requiring the propagated proton spectrum to be in agreement with the local proton measurements by AMS-01 \cite{AMS}, BESS \cite{Sanuki:2000wh}, and IMAX \cite{2000ApJ...533..281M} above 1 GeV, and the propagated antiproton spectrum and the antiproton-over-proton ratio $\bar{p}/p$ to be in agreement with the recent PAMELA measurements \cite{Adriani:2010rc,PAMELA:2010rc}.  Accordingly, for the protons we take the power law index below $E_b=9$ GeV to be $\gamma_{1}=-1.98$.  We note that our fits to the PAMELA and \emph{Fermi} \epm data do not extend below 10 GeV, so the spectrum of protons below 10 GeV does not affect our fits.  Therefore, we do not vary the form of the proton spectrum below the break energy.  Above 9 GeV, we allow the power law index to vary such that the resulting spectrum does not exceed the local data by more than $3 \sigma$.  The PAMELA $\bar{p}/p$ data place the strongest constraint on the range of acceptable power law indices for the injection spectrum of primary protons above 9 GeV.  We find the allowed range to be $-2.50 \le \gamma_{2} \le -2.10$.

We use the \emph{Fermi} \cite{Abdo:2009zk} electron data to constrain the injection spectrum of primary electrons above 10 GeV and confirm our findings with several other measurements of the electron spectrum \cite{Boezio:2000,DuVernois:2001bb,Torii:2001aw,Grimani:2000yz}.  Fitting to the \emph{Fermi} data between $\sim 70$ GeV and $\sim 380$ GeV, we find that the range of indices $-2.57$ to $-2.18$ gives spectra that are consistent with the data such that no \emph{Fermi} data point is exceeded by more than $3 \sigma$.  We stress that these constraints arise from fitting only the primary electrons to the $\epm$ \emph{Fermi} data; we have not included the fluxes of secondary electrons and positrons in our fits.  Since the secondaries are an order of magnitude or more smaller than the flux of primary $e^-$'s between 10 GeV and 100 GeV, including them would result in only a small softening of the allowed spectrum, making the range of indices slightly more negative.  However, inclusion of a dark matter component of $\epm$ in the flux allows for considerable softening of the primary electron spectrum.  To accommodate the effects of a large dark matter \epm component, we extend the range of allowed indices to smaller values and perform our scans over the range $-3.00$ to $-2.00$.  Our fits to the PAMELA and \emph{Fermi} data confirm that this is a reasonable choice, as only a few of the $2 \sigma$ \emph{Fermi} fits require an electron injection spectrum with an index smaller than $-3.00$ and none of the fits require an index larger than $-2.40$.  The H.E.S.S data indicate a steep falloff in the $\epm$ spectrum above $\sim 1$ TeV \cite{Collaboration:2008aaa, Aharonian:2009ah}.  We assume that the local primary electron spectrum exhibits this behavior and introduce a break at 2.2 TeV, above which the injection spectrum has a power law index of $-3.30$.

\subsection{Propagation parameters}
\label{sec:propparams}
We define the cylindrical diffusion zone centered on the Galactic Plane (GP) to have a radius of $r_d=20\kpc$ and height $h_d=\pm4\kpc$.  While one may find reference to diffusion zone heights as small as $h_d=1\kpc$ and as large as $h_d=15\kpc$, these are extreme values.  Standard values lie in the range $3-7$ kpc.  Inside the diffusion zone, cosmic ray propagation is governed by the diffusion-loss equation.  At the boundaries of the diffusion zone, free escape of all cosmic rays is assumed to occur.  Measurements of the local value of the unstable secondary-to-primary ratio $\rm Be^{10}/Be^{9}$ provide a good check on the choice of diffusion zone height, since this ratio of secondaries-to-primaries is dependent on the average time cosmic rays spend in the Galaxy before free-escape, decay, and spallation.  The energy dependent diffusion coefficient has the form $D(E)=D_0\times10^{28}\; (E/4\GeV)^\alpha \cm^2 \s^{-1}$.  Typical values of $\alpha$ are between 0.33 and 0.50 (with $\alpha=0.33$ corresponding to a turbulent magnetic field with a Kolmogorov type spectrum and $\alpha=0.50$ corresponding to a turbulent field with a Kraichnan type spectrum), while $D_0$ takes values in the range $\sim2 - 10$.  The stable secondary-to-primary ratios B/C and sub-Fe/Fe are a measure of the average amount of matter cosmic rays propagate through.  For this reason, measurements of B/C and sub-Fe/Fe help constrain the diffusion coefficient.  We calculate our fits to the PAMELA and \emph{Fermi} data using the benchmark diffusion parameters $D_0 = 4.00$ and $\alpha = 0.50$, which are broadly similar to the M1 parameters of \cite{Delahaye:2007fr}, albeit with a more conventional disk scale height (i.e. $h_d = 4$kpc rather than the 15 kpc of the M1 model), and fit the data for these three secondary-to-primary ratios at the $3 \sigma$ level. The effect on the local cosmic-ray spectra of variations in the diffusion parameters can largely be compensated for by changes in the background spectral indices, so we choose to fix the diffusion parameters and vary the background as described above.

We model the magnetic field as an exponential disk, 
\be
\langle B^2\rangle^{1/2} = B_0 \exp \biggl[-\frac{(r-R_{\odot})}{r_B} - \frac{|z|}{z_B}\biggr],
\ee
where $r_B=10\kpc$ is the radial scale, $z_B=2\kpc$ is the vertical scale, and $B_0 = 5.0\, \mu G$ is the local value.  Additionally, we use the most current model of the interstellar radiation field available to the public, that of \cite{Porter:2005qx}.

\subsection{Dark matter density}
For our choice of dark matter density profile, we use an Einasto profile \cite{Einasto,Merritt:2005}
\begin{equation}
\label{eq:SHM_eq}
	\rho(r)=\rho_{0} \exp\left[-\frac{2}{\alpha}\left(\frac{r^{\alpha}-R_{\odot}^{\alpha}}{r_{-2}^{\alpha}}\right)\right],
\end{equation}
with $\alpha=0.17$ and $r_{-2}=25$ kpc. The local density is uncertain, but the best current estimates are $\rho_0 /(\mathrm{GeV} \, \mathrm{cm}^{-3}) = 0.385\pm 0.027$ \cite{Catena:2009mf}, $0.43 \pm 0.11 \pm 0.10$ \cite{Salucci:2010qr}, and $0.466 \pm 0.033 \pm 0.077$ \cite{Pato:2010yq}. We conservatively take $\rho_0=0.4$ GeV/cm$^3$, but note that taking a $\sim 1 \sigma$ high value from the result of \cite{Pato:2010yq} would lower the needed boost by a factor of 2. Since the majority of the \epm signal is produced in the nearby $\sim 1$ kpc, the question of the dark matter density profile -- Einasto, NFW, cored isothermal or something else -- is not particularly important.

\subsection{Quality of fits}
\label{sec:qualityoffits}
Others have previously employed $\chi^2$ analyses to determine the best-fit regions for particular final states \cite{Cirelli:2008pk,Meade:2009rb,Meade:2009iu}; however, in the absence of a covariance matrix for the data, the information that can be extracted from the $\chi^2$ is limited. The usual assumption is that the error bars are completely uncorrelated, and this is indeed a conservative approach for the simple question of whether a given model is a good fit ($\chi^2$/dof $< 1 + \epsilon$). However, the question of whether one model with $\chi^2 \lesssim 1$ is a better fit than another, and with what confidence, cannot be answered without the covariance matrix, since if one simply assumes uncorrelated errors then the data is skewed by repeatedly counting correlated errors, erroneously disfavoring regions of parameter space in an unpredictable way. 

As an example, imagine we have a dataset with perfectly correlated large systematic errors ($\sim \sigma$), and very small statistical errors ($\sim \epsilon$). Suppose we have two models, one of which is consistently $\sim 1 \sigma$ above the data, and one of which varies smoothly from $\sim 1 \sigma$ above the data at the lowest energies to $\sim 1 \sigma$ below the data at the highest energies. If we incorrectly assume that all the errors are uncorrelated, and add the systematic and statistical errors in quadrature, then the $\chi^2$/dof will be $\sim 1$ for the first model and $\sim 1/3$ for the second model; if the number of degrees of freedom is large, a naive $\Delta \chi^2$ analysis will find that the second model is preferred at very high confidence. If we take into account the perfect correlation of the systematic errors, however, then the first model still has $\chi^2/\mathrm{dof} \approx 1$, but the $\chi^2$/dof for the second model balloons by a factor of $\sim (\sigma/\epsilon)^2$. The conservative approach of taking all the errors to be uncorrelated will never rule out a model that is actually a good fit, but a model that appears to be the best fit may no longer be the best fit, and in fact may be ruled out, once the full covariance matrix is taken into account.

Consequently, to determine if a particular choice of $(m_\chi, m_\phi, \delta)$ fits the data, we insist that the background+signal curves pass through the error bars for the PAMELA positron fraction data above 10 GeV and the \emph{Fermi} $e^+ + e^-$ data above 30 GeV (where charge-dependent and charge-independent solar modulation, respectively, are thought to be small), for \emph{some} set of parameters governing the background electron and positron spectra, within the bounds described in  \S \ref{sec:BGspectra}.  We compute the regions where these conditions are satisfied for the $1 \sigma$, 90\% confidence, and $2 \sigma$ error bars.  We additionally require that the curves not exceed the H.E.S.S $e^+ + e^-$ systematic error band, which is, in effect, forcing the $e^+ + e^-$ spectrum to satisfy the $E^{-4}$ fall-off above $\sim 1$ TeV.  Using the $1 \sigma$ errors, these requirements are more constraining than demanding $\chi^2$/dof $< 1 + \epsilon$, allowing us to select a preferred region in the allowed parameter space without having to explicitly write down a model for the covariance matrix. This approach has the positive features that:
\begin{itemize}
\item It disfavors models where a small number of points lie well outside the error bars; since the true uncorrelated fluctuations are likely to be much smaller than the nominal error bars, this is a positive feature.
\item It treats different datasets on an equal footing; the favored parameters provide model curves in good agreement with PAMELA \emph{and} \emph{Fermi}. In contrast, with a $\chi^2$ test assuming uncorrelated errors the ``best-fit'' models tend to have $\chi^2$/dof $\ll 1$ for the \emph{Fermi} data, which compensates for a somewhat poor fit to the PAMELA data. This skews the ``best-fit'' regions towards lower values of the boost factor.
\end{itemize}
This method makes no claims to determine the relative likelihoods of models that each fit the data well. In the absence of covariance matrices from the PAMELA and \emph{Fermi} experiments, it is probably the most reasonable quantitative test of ``explaining the data'' that is available to us. 

With regard to the H.E.S.S data, two points are relevant here. First, we follow \cite{Bergstrom:2009fa} and treat the H.E.S.S data as an upper limit. Although photon contamination is expected to be low at these energies, the data are still properly limits. Second, as noted by \cite{Bergstrom:2009fa}, the data themselves seem inconsistent with the \emph{Fermi} data. Attempting to fit {\em both} is then likely to be partially fitting to a systematic uncertainty. Consequently, we do not fit the H.E.S.S data, but do show them for our benchmark points, rescaled to be consistent with \emph{Fermi}. We differ slightly from \cite{Bergstrom:2009fa} here, in that we assume that a small ($\sim$ 8\%) shift in the energy scale explains this, rather than a 15\% shift in overall normalization. While these data are not explicitly included in our fits, our benchmark points nonetheless agree well with the rescaled H.E.S.S data.

As a check on our method, we also computed the $\chi^2$ as a function of $m_\mathrm{DM}$ and BF, choosing the best-fit background parameters at each point. The allowed regions -- that is, the regions that are \emph{not} excluded by a goodness-of-fit $\chi^2$ test, at 90$\%$, $95\%$ and $99\%$ confidence -- are quite large, and as mentioned above, they tend to skew toward lower boost factors due to the large number of \emph{Fermi} data points (with covariances that are not taken into account). Inclusion of the rescaled H.E.S.S data in the $\chi^2$ (instead of using H.E.S.S only to set upper limits) substantially reduces the size of the allowed regions; however, all of our benchmark points, and the bulk of our preferred parameter space, remain allowed. Some small regions with high boost factors and relatively low DM masses are ostensibly ruled out: this is because such models require a fairly soft background electron spectrum, to fit the data with a large DM contribution, and at energies above the mass of the DM this soft background spectrum undershoots the H.E.S.S data. It is possible that by varying the position of the high-energy break in the background electron spectrum, or allowing a small shift in the overall energy scale, a better fit might be achievable even for these models, but in any case, the effect of a conservative $\chi^2$ analysis is generally to shift the preferred range of boost factors \emph{downward}.

\section{Results}
\label{sec:results}

\subsection{Parameter scans}
\label{sec:plotssec}

To determine the current annihilation rate, we scan a three-dimensional parameter space consisting of the DM mass $m_\chi$, the dark gauge boson mass $m_\phi$, and the mass splitting $\delta$. We focus on the singly-charged Higgs model with the mass splitting generated as described in \S \ref{sec:masssplittingann}, which we refer to as the ``minimal model''. The dark sector coupling $\alpha_D$ is determined by requiring the thermal relic density to be $\Omega h^2 = 0.1120$, in accordance with the WMAP7+BAO+H0 ML cosmological parameter set \cite{Komatsu:2010fb,Percival:2009xn,2009ApJ...699..539R}. The fits to the PAMELA and \emph{Fermi} cosmic-ray excesses are determined by $m_\chi$, which sets the overall energy scale, and $m_\phi$, which determines the branching ratios of the dark gauge boson to different SM final states (as summarized in e.g. \cite{Falkowski:2010cm}).

\begin{figure}[p]
\centering
\begin{tabular}{cc}
\includegraphics[width=0.45\textwidth]{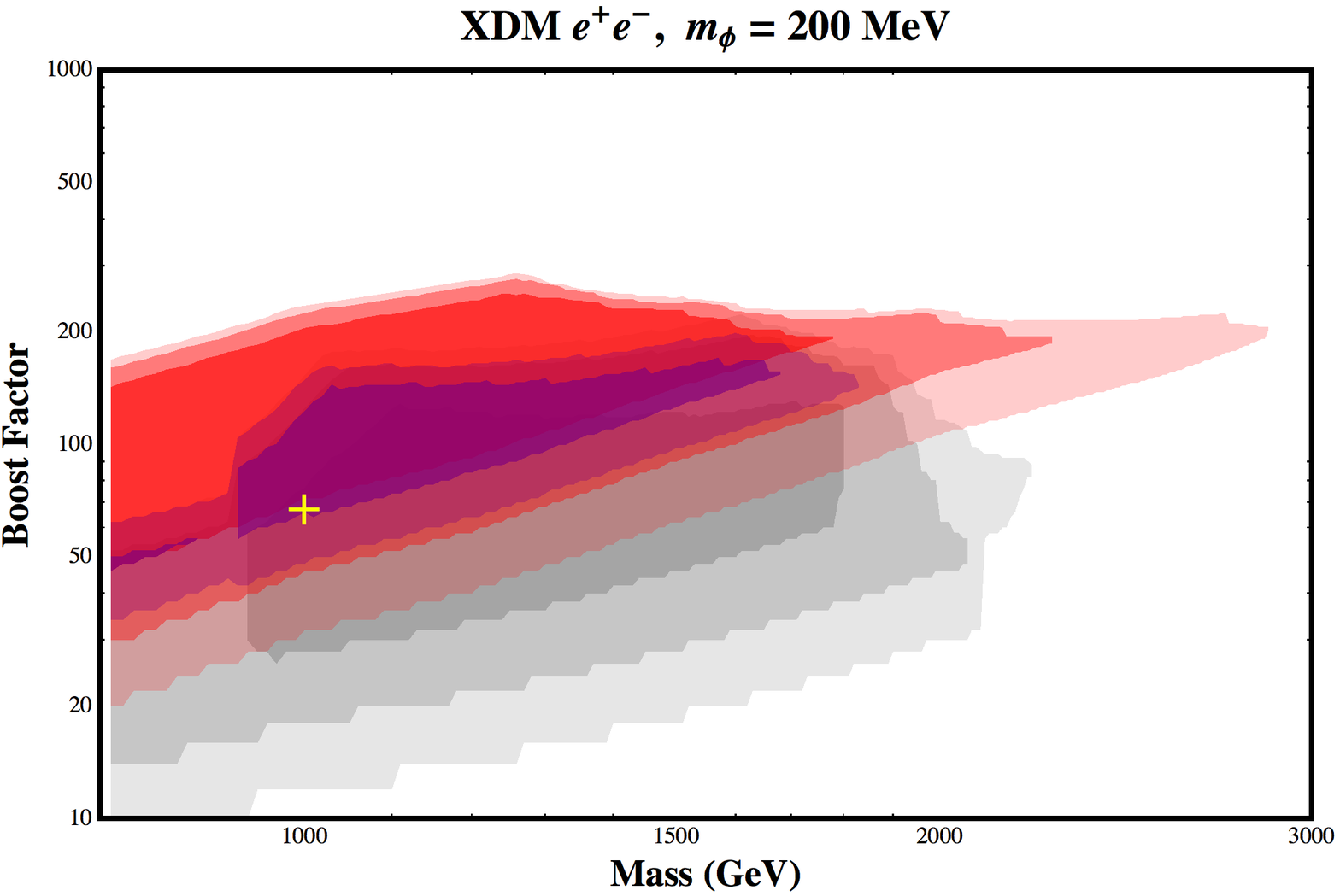} &
\includegraphics[width=0.45\textwidth]{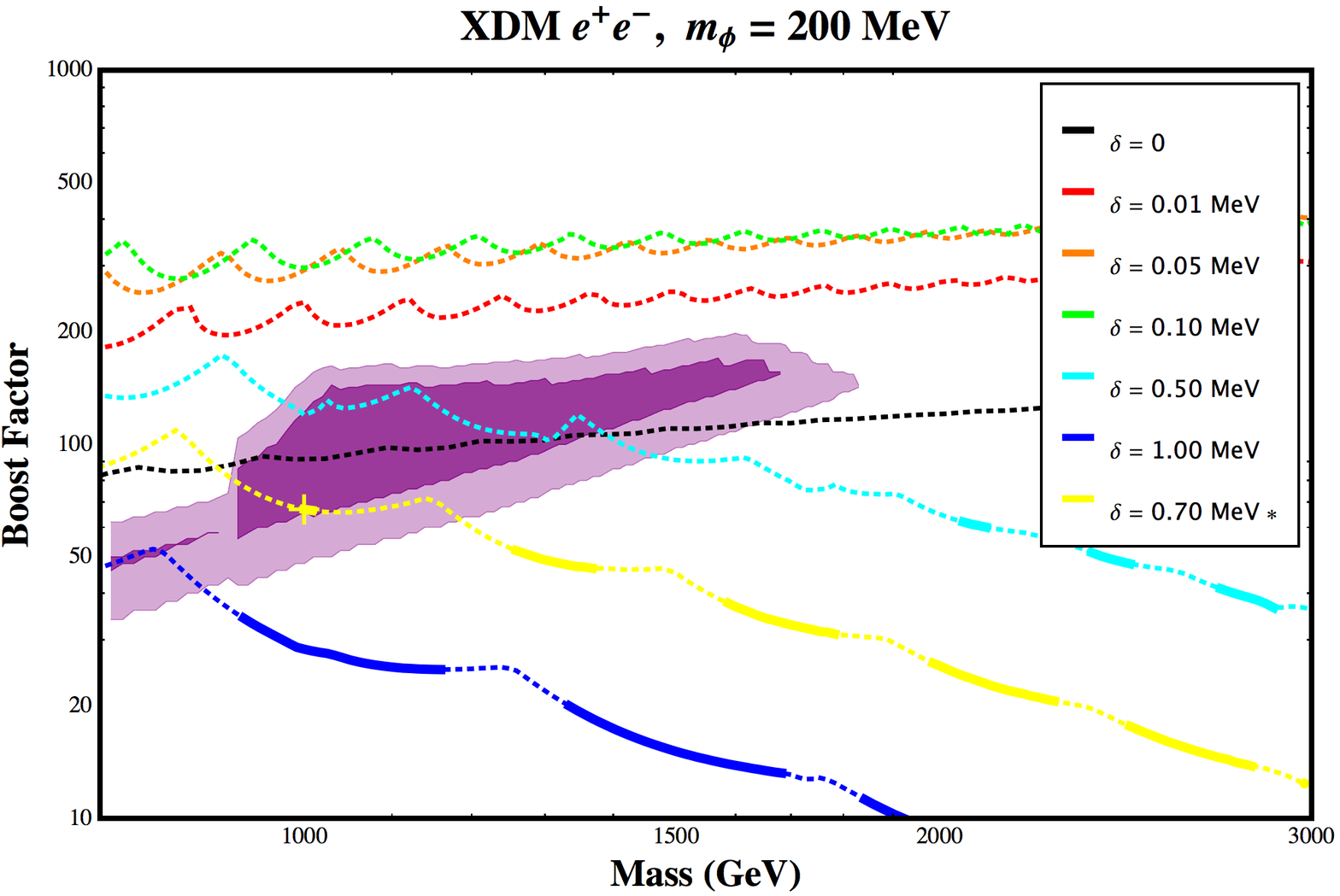} \\
\vspace{1pt}
\includegraphics[width=0.45\textwidth]{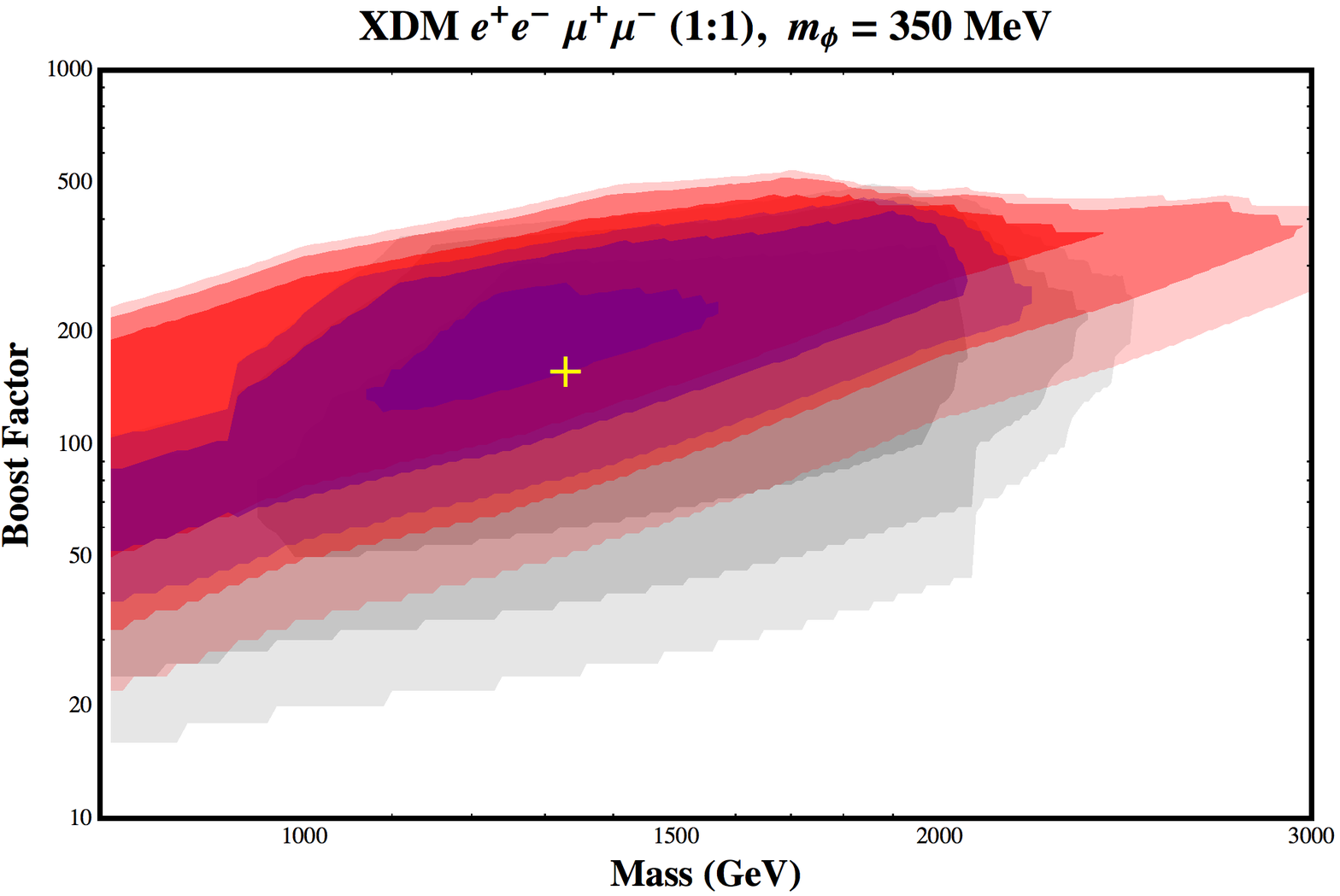} &
\includegraphics[width=0.45\textwidth]{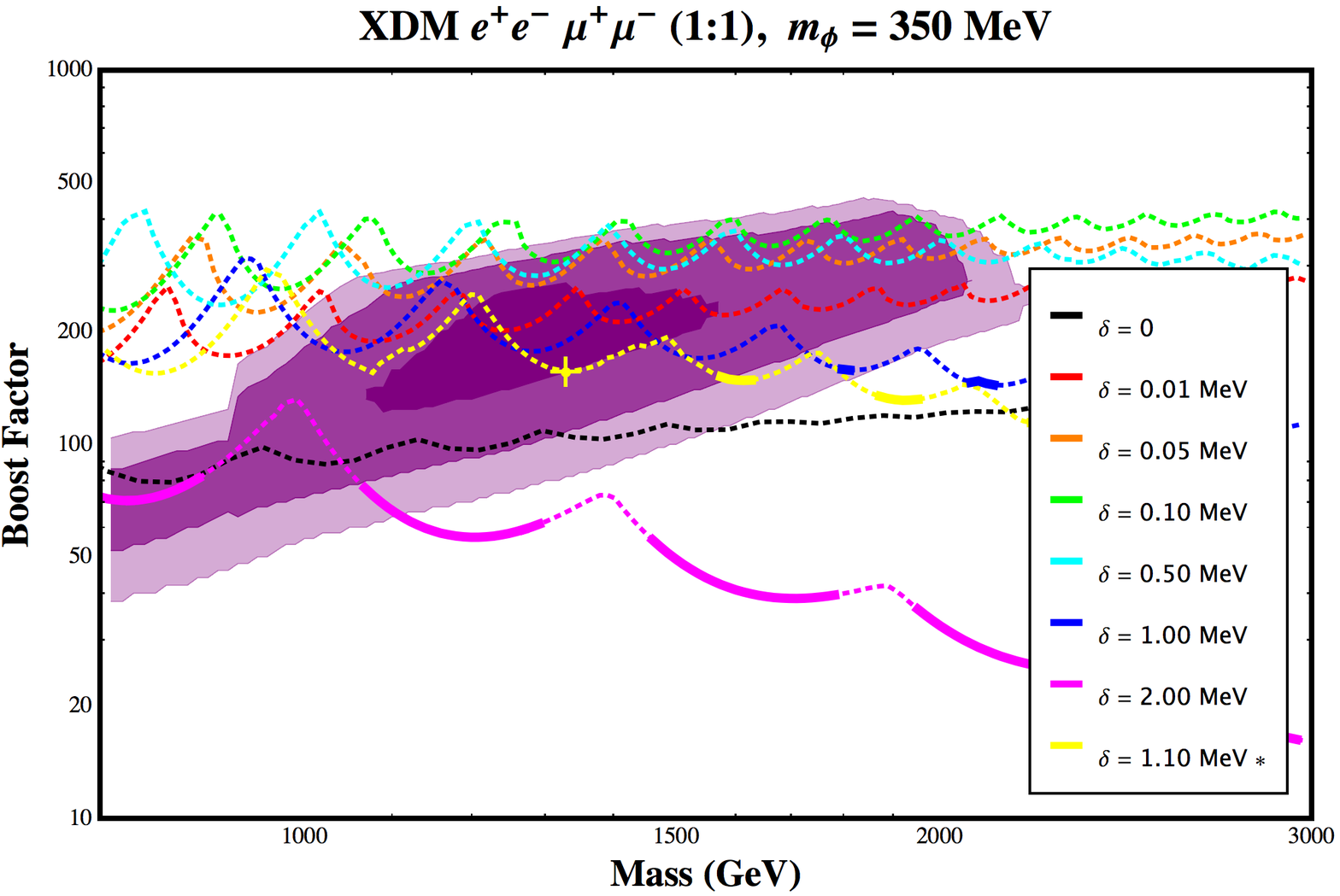} \\
\vspace{1pt}
\includegraphics[width=0.45\textwidth]{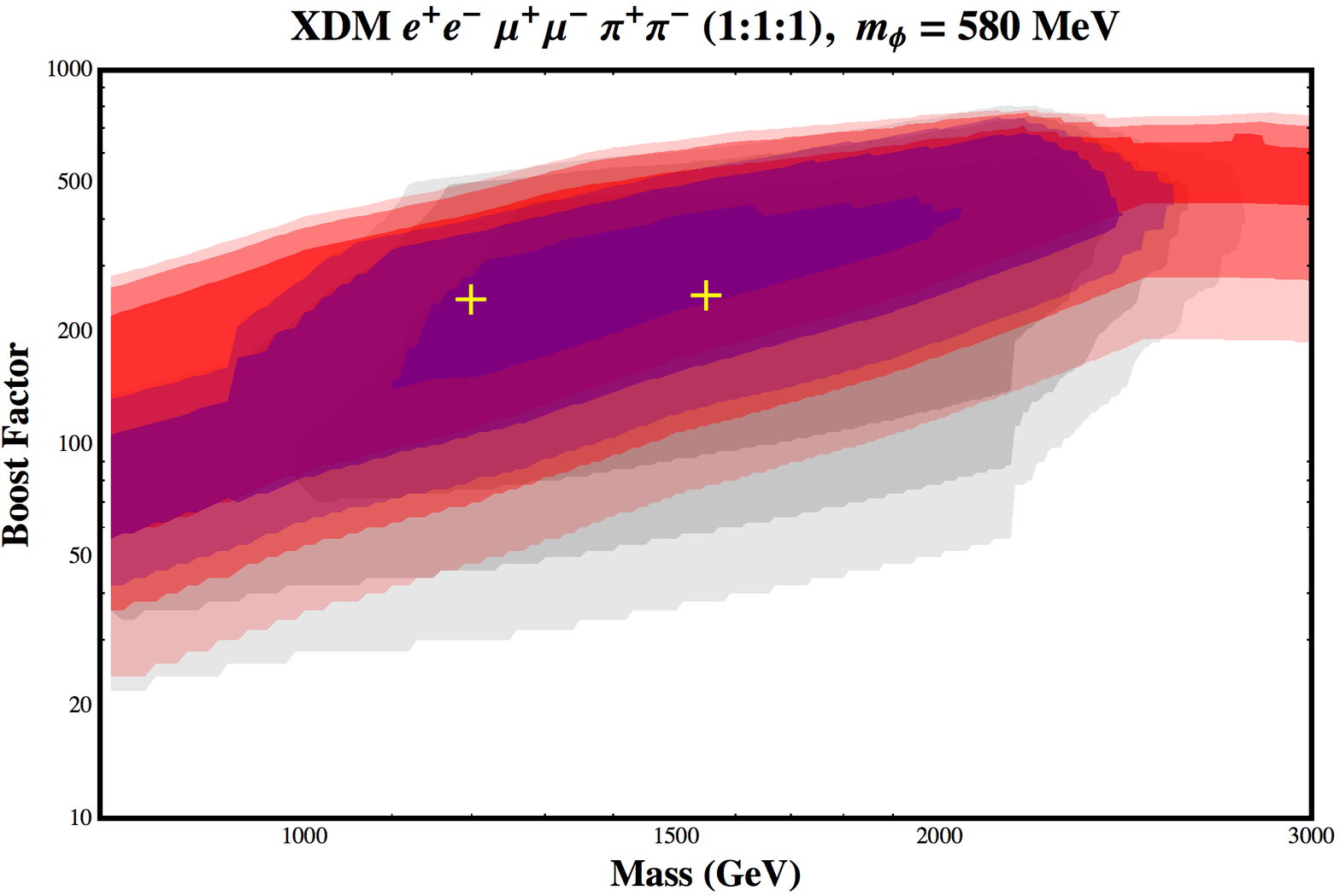} &
\includegraphics[width=0.45\textwidth]{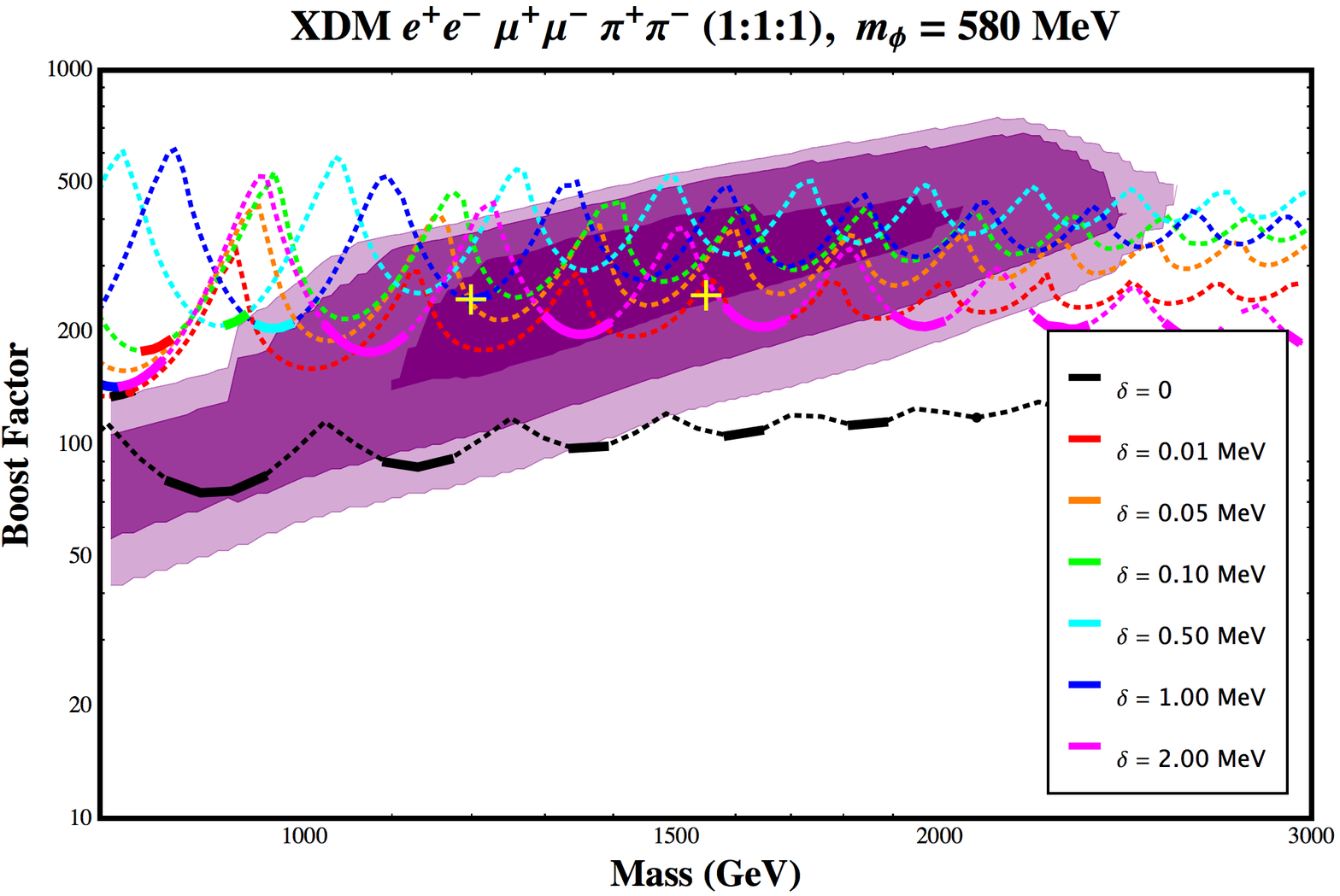} \\
\vspace{1pt}
\includegraphics[width=0.45\textwidth]{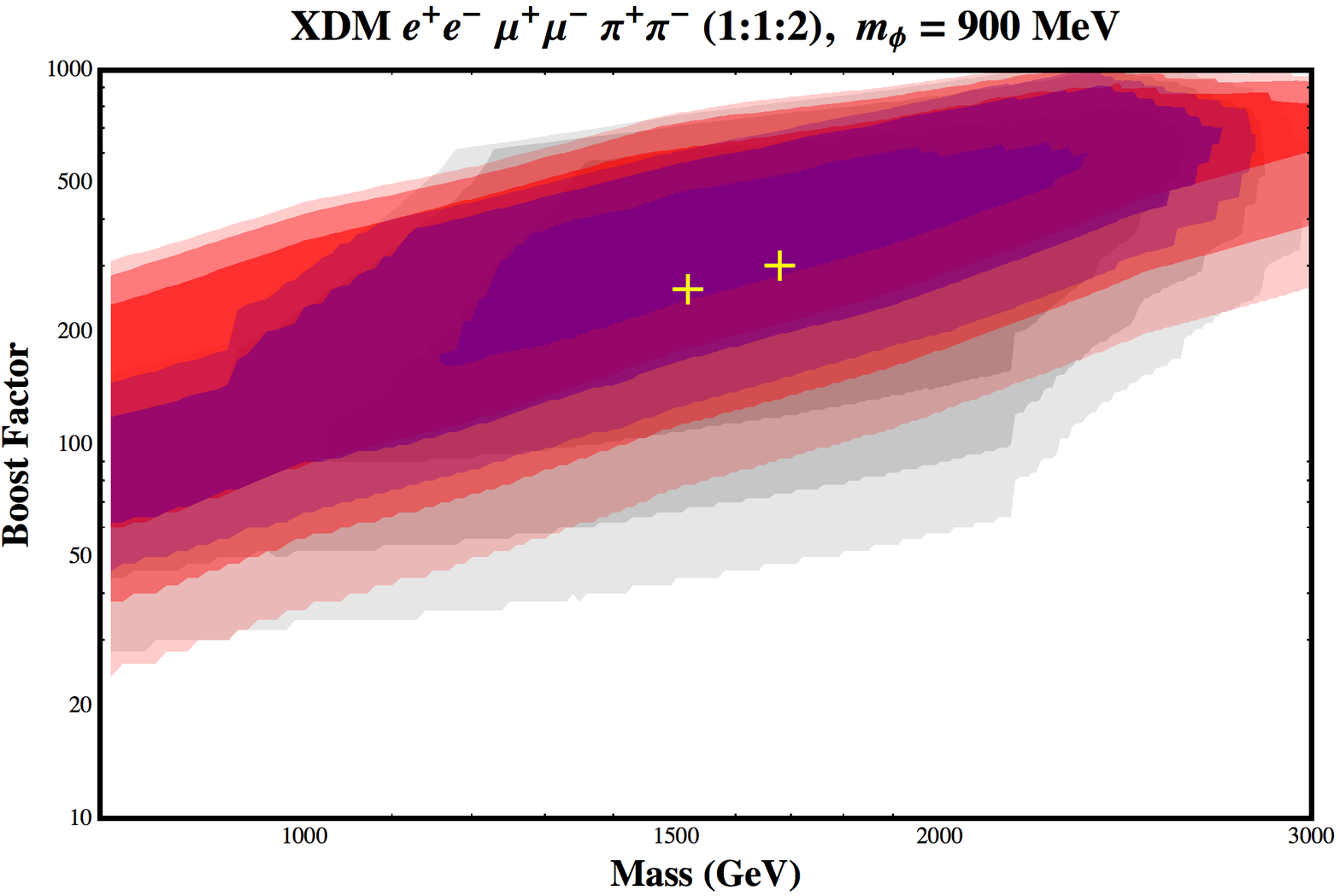} &
\includegraphics[width=0.45\textwidth]{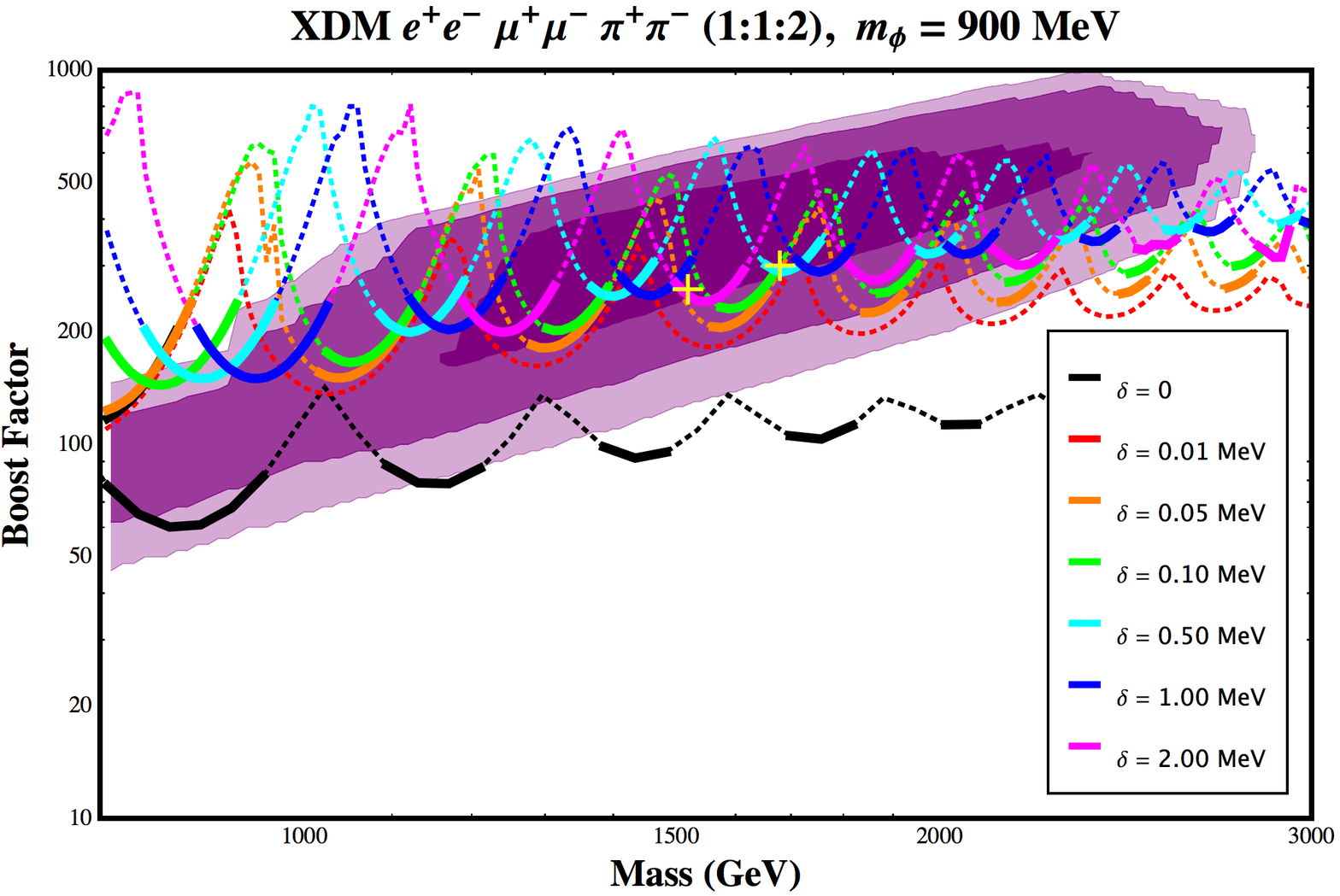} \\
\end{tabular}
\caption{\footnotesize \emph{Left:} Allowed ranges of parameter space for fits within the $1 \sigma$, 90\% confidence, and $2 \sigma$ error bars to PAMELA only (in decreasing intensity of \emph{red}), \emph{Fermi} only (in decreasing intensity of \emph{gray}), and for simultaneous fits to both PAMELA and \emph{Fermi} (in decreasing intensity of \emph{purple}). Yellow crosses indicate benchmark points. \emph{Right:} As in \emph{left}, with curves showing the boost factors for a range of mass splittings $\delta$ such that $\Omega h^2 = 0.1120$ (\emph{dashed}). Yellow lines, marked with asterisks, are chosen to pass through the benchmark points for cases where the BF varies rapidly with $\delta$.  The CMB constraints are met for the \emph{solid} portions of the curves.  Results are shown for 800 GeV $\le m_{\chi} \le$ 3 TeV only. All preferred regions shown here assume $\rho_0 = 0.4$ GeV/cm$^3$ and \emph{no} contribution to the signal from DM substructure; any substructure correction (e.g. \cite{Pieri:2009je}) will shift the preferred regions to lower boost factors. The $\delta=0$ curve is intended as a consistency check with previous work, and so annihilation channels involving the dark Higgs were omitted from the computation in this case.}
\label{fig:allowedparams}
\end{figure}

We show in Figure \ref{fig:allowedparams} the allowed BF-mass parameter space for four representative final states, corresponding to four benchmark mediator masses: $\phi \rightarrow \epm$ ($m_\phi = 200$ MeV), $\phi \rightarrow \epm$:$\phi \rightarrow \mu^+\mu^-$=1:1 ($m_\phi = 350$ MeV), $\phi \rightarrow \epm$:$\phi \rightarrow \mu^+\mu^-$:$\phi \rightarrow \pi^+\pi^-$=1:1:1 ($m_\phi = 580$ MeV), and $\phi \rightarrow \epm$:$\phi \rightarrow \mu^+\mu^-$:$\phi \rightarrow \pi^+\pi^-$=1:1:2 ($m_\phi = 900$ MeV). At higher mediator masses, hadronic final states become increasingly important and the softness of the spectrum makes it more challenging to fit the data.  In the left-hand column of Figure \ref{fig:allowedparams} we show the fits lying within the $1 \sigma$, 90\% confidence, and $2 \sigma$ error bars for PAMELA only, for \emph{Fermi} only, and for \emph{both} PAMELA and \emph{Fermi}.  In the right-hand column we again show the fits within the $1 \sigma$, 90\% confidence, and $2 \sigma$ error bars for both PAMELA and \emph{Fermi}.  

We briefly clarify some of the assumptions made in determining the parameter space shown in Figure \ref{fig:allowedparams}.  As discussed in \S \ref{sec:BGspectra} and \S \ref{sec:propparams}, we describe the background proton spectrum as a power law with one break.  For our analysis, we fix both the break energy and the index below this break energy.  See Table~\ref{tab:benchmarks} for values.  We allow the upper index $\gamma_{p2}$ to vary in the range $(-2.50, -2.10)$ and the overall proton normalization to vary.  Similarly, we describe the electron spectrum as a power law with two breaks.  We hold the spectral index fixed for energies below $E_{e1}=4$ GeV.  Moreover, the high-energy index, $\gamma_{e3}=-3.30$, and break energy, $E_{e2}=2200$ GeV, are held fixed at values chosen to give agreement with the \emph{Fermi} and H.E.S.S data at energies above $\sim 1$ TeV.  We are left with two varying parameters for the electron spectrum, the electron spectral index in the energy range 4 to 2200 GeV, which is allowed to vary between $-3.00$ and $-2.00$, and the overall normalization of the spectrum.  In summary, the BF-mass parameter space shown in Figure \ref{fig:allowedparams} is determined by varying four parameters, the proton spectral index above 9 GeV (which can take values in the range $(-2.50, -2.10)$), the electron spectral index between 4-2200 GeV (which can take values in the range $(-3.00, -2.00)$), and the normalization of each spectrum.  As discussed in \S \ref{sec:propparams}, the effects of the diffusion parameters on the background cosmic-ray spectra can be mimicked by a change in the spectral indices.  Thus, a variety of diffusion scenarios are represented in the analysis leading to the results shown in Figure \ref{fig:allowedparams}. A BF-mass combination is considered to fit the data if, within these constraints, the background parameters can be chosen so the signal+background curve passes through every error bar (for example, the ``1$\sigma$'' regions require the model curve to pass through every 1$\sigma$ error bar).

We include curves in Figure \ref{fig:allowedparams} describing the allowed values of the local boost factor (BF) for several different DM mass splittings $\delta$, subject to the relic density constraint. Recall that the boost factor is defined as $\langle \sigma v \rangle / 3 \times 10^{-26} \rm cm^3 \;s^{-1}$. The solid portions of the curves are those for which the CMB constraints are met.  In both sets of plots we mark with a ``+" benchmark points for each mediator mass; a benchmark point is one that simultaneously fits the PAMELA and \emph{Fermi} data with the correct relic density, while satisfying the CMB constraints.  Results are shown for 800 GeV $\le m_{\chi} \le$ 3 TeV only, though the allowed regions may extend to lower and higher masses. For comparison to earlier work (e.g. \cite{Zavala:2009mi, Feng:2010zp}), we also show the case with no mass splitting; in this case we omit annihilation channels involving the dark Higgs, since such annihilation channels were not included in the previous studies. It is clear that the introduction of an $\mathcal{O}$(MeV) mass splitting removes any tension between the preferred PAMELA/\emph{Fermi} region and the boost factor from Sommerfeld enhancement. However, the CMB constraints introduce significant tension at lower mediator masses.

\afterpage{\clearpage}

\begin{figure}[t]
\centering
\includegraphics[width=0.3\textwidth]{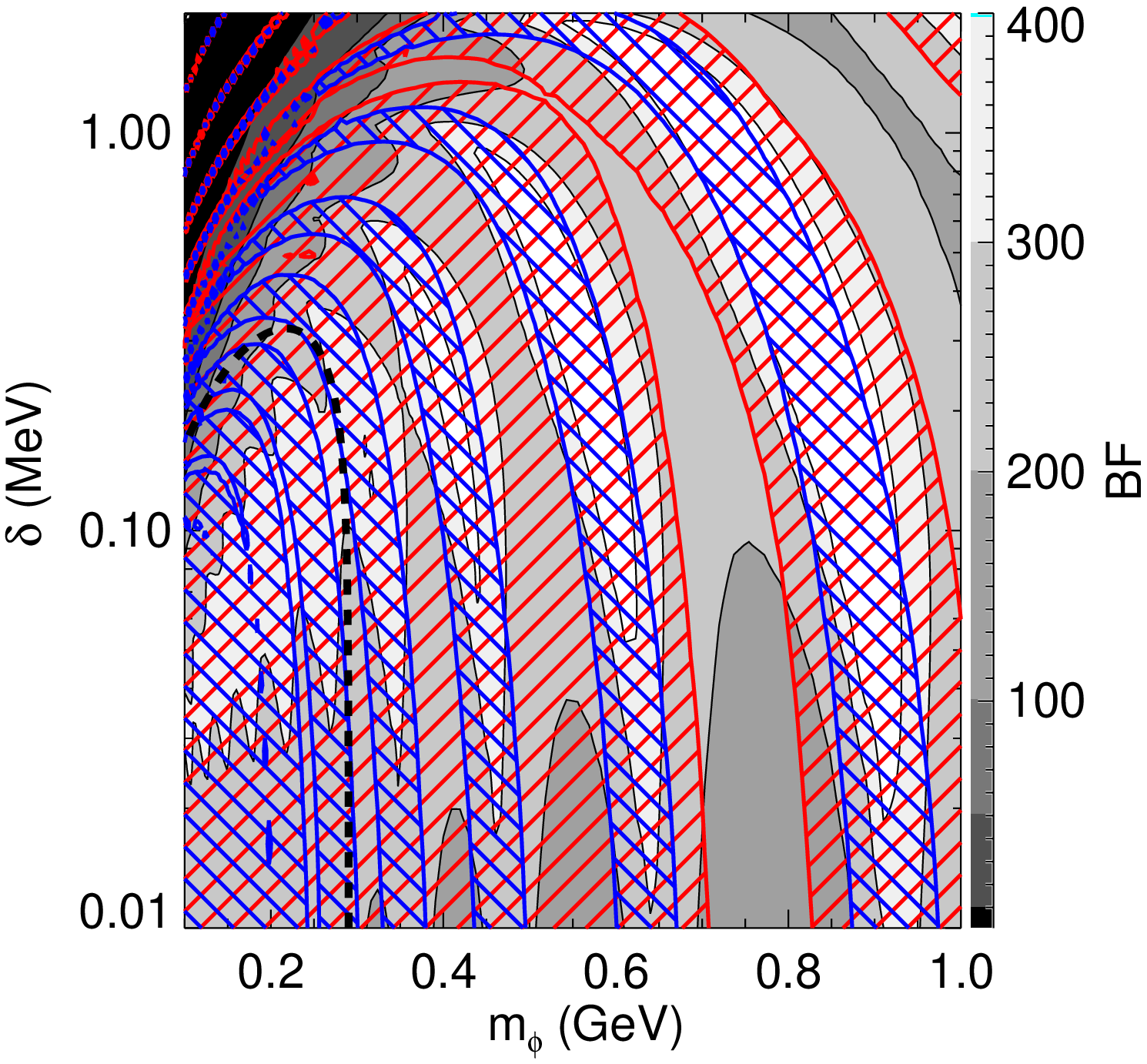}
\includegraphics[width=0.3\textwidth]{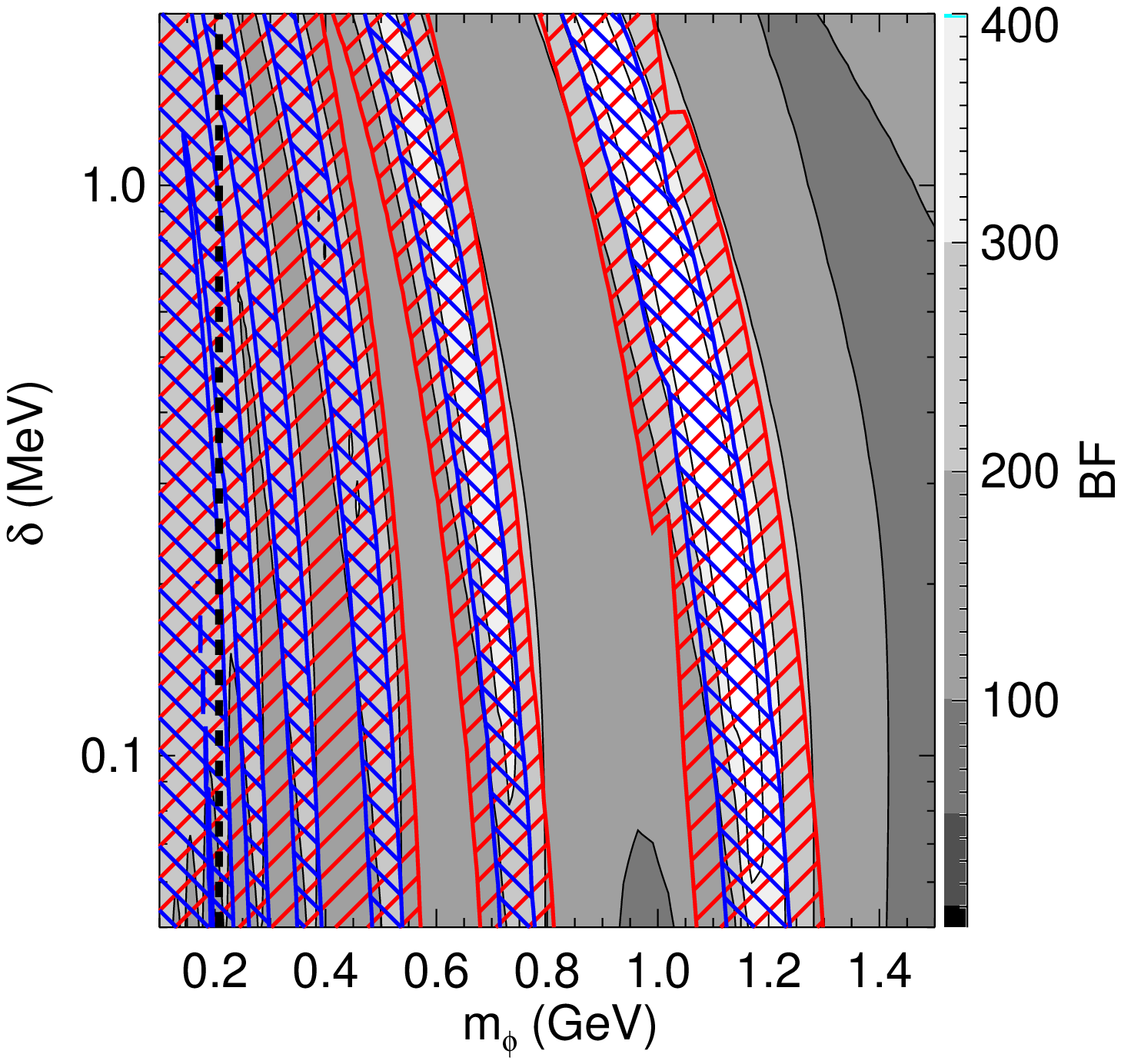} 
\includegraphics[width=0.3\textwidth]{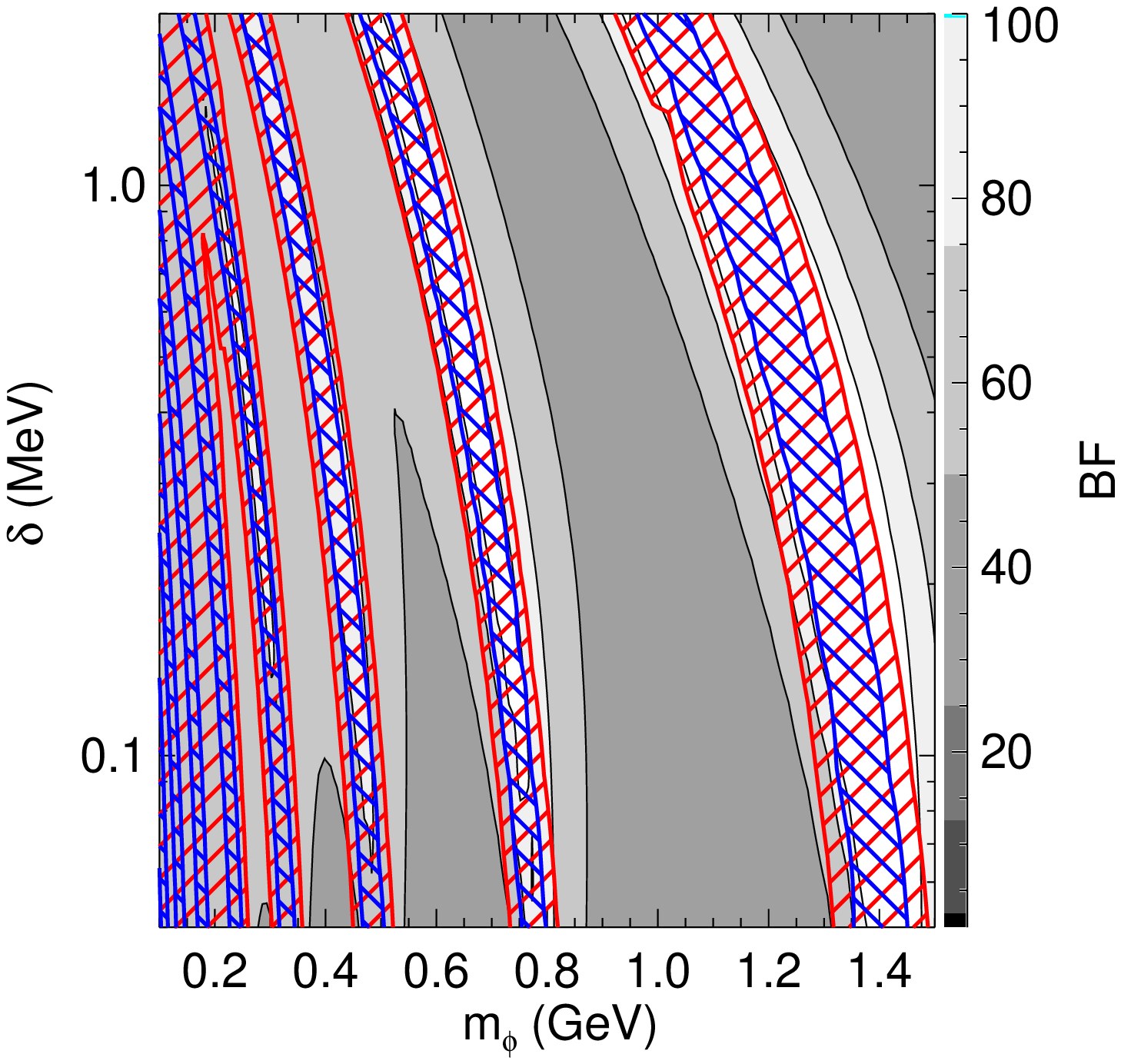} \\
\includegraphics[width=0.3\textwidth]{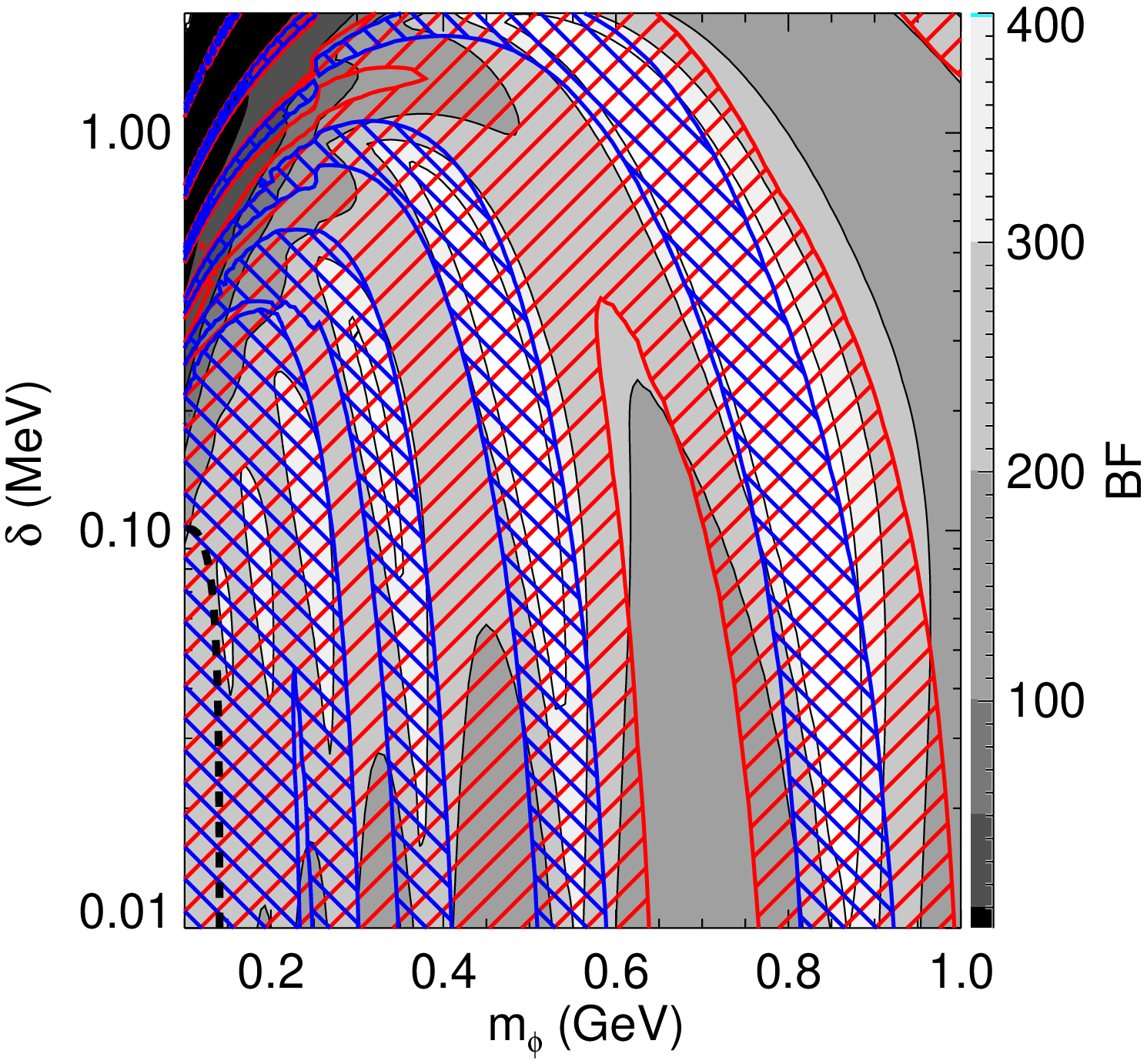}
\includegraphics[width=0.3\textwidth]{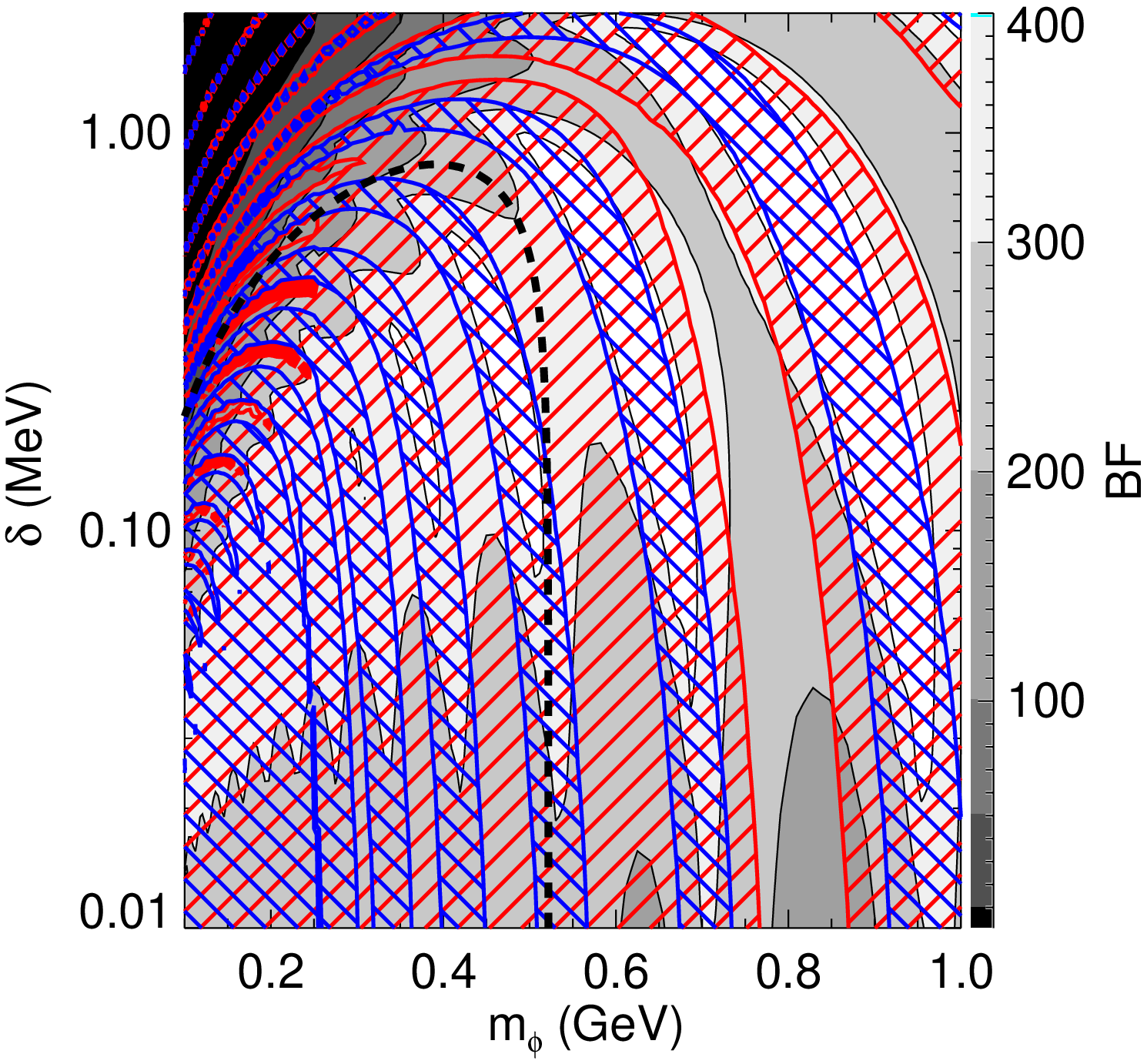}
\includegraphics[width=0.3\textwidth]{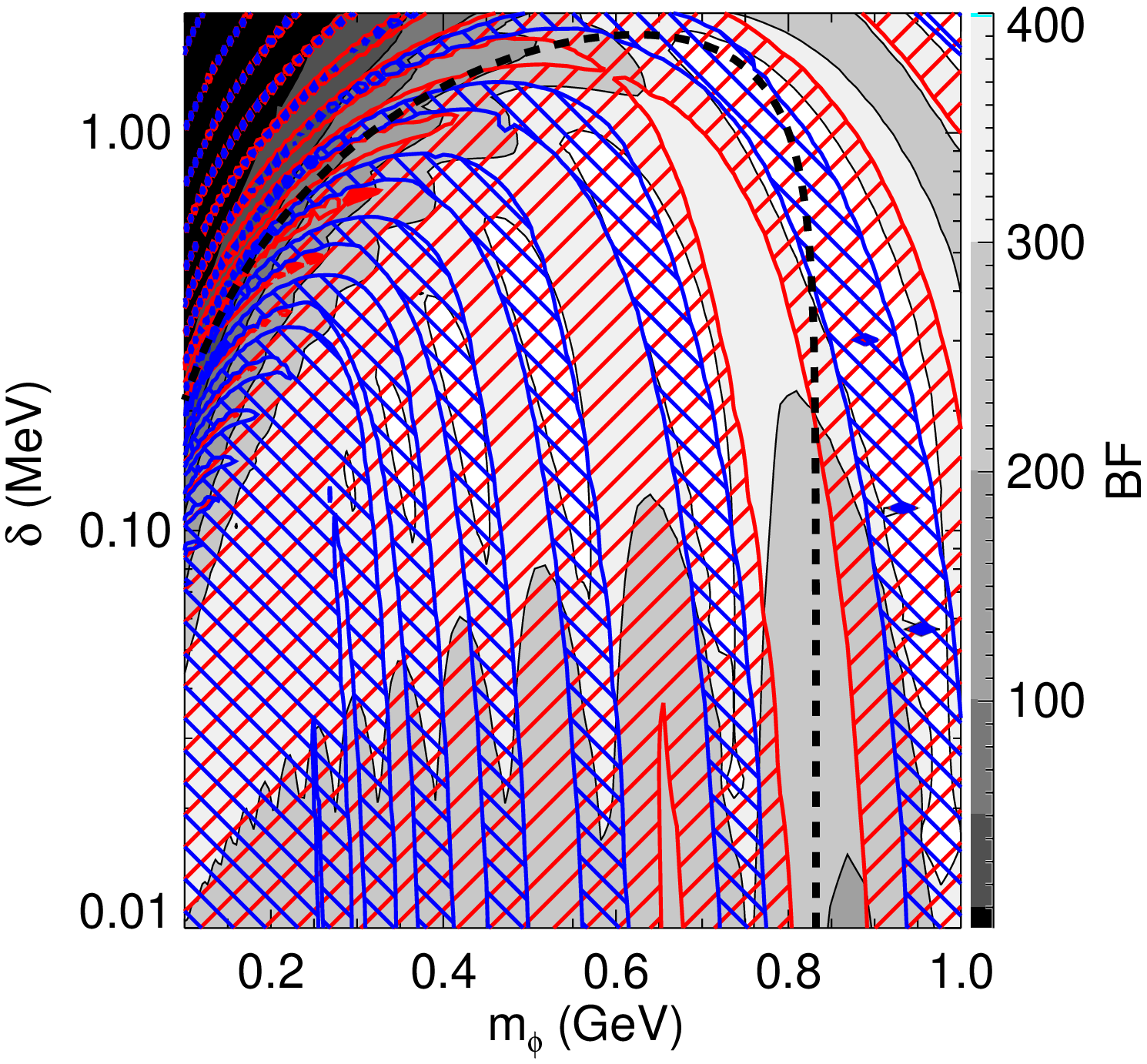}
\caption{\footnotesize Contours for the boost factor BF in the local halo as a function of the mediator mass $m_{\phi}$ and mass splitting $\delta$, with $\alpha_D$ chosen to produce the correct relic density. The DM velocity distribution is assumed to be Maxwellian with $\sigma = 150$ km/s. The regions overlaid with red lines are ruled out at $95 \%$ confidence by bounds from WMAP5, taking $f = 0.2$ for hadronic final states, $f = 0.24$ for muons, and $f = 0.7$ for electrons, and weighting these three contributions according to the ($m_\phi$-dependent) branching ratios for decays of the dark gauge boson. Blue-hatched regions illustrate the effect of shifting the WMAP5 constraint: these parts of parameter space would remain ruled out even if the bounds were relaxed by a factor of 4 (e.g. due to degeneracy with some parameter not included in the analysis, although we do not consider this a likely scenario). \emph{Upper row:} Results for 1.2 TeV dark matter, for three values of the coannihilation parameter $\kappa$.  \emph{Upper left panel:} $\kappa = 1/4$, corresponding to the minimal singly-charged Higgs model described in \S \ref{sec:masssplittingann}.  \emph{Upper center panel:} $\kappa = 1$ ($\alpha_D = 0.0263$, see text). \emph{Upper right panel:} $\kappa = 4$ ($\alpha_D = 0.0177$, see text). \emph{Lower row:} Results for DM mass (\emph{lower left panel}) 900 GeV, (\emph{lower center panel}) 1.5 TeV, and (\emph{lower right panel}) 1.8 TeV, in the $\kappa=1/4$ minimal model.  Capture into a bound state, inducing an additional enhancement to late-time annihilation, is kinematically allowed in regions to the left of and below the black-dashed line, but has \emph{not} been included in the analysis; see Appendix \ref{sec:wimponium} for a discussion.}
\label{fig:paramscan}
\end{figure}

In Figure \ref{fig:paramscan}, we hold the DM mass fixed and scan over $m_\phi$ and $\delta$, for $m_\chi=0.9,1.2,.1.5$ and 1.8 TeV in the minimal model. For illustration, for $m_\chi=1.2$ TeV we also show the effect of changing the coannihilation parameter to $\kappa=1,4$. For $\kappa > 1/4$ we include only the self-annihilation channel $\chi \chi \rightarrow \phi \phi$, and a coannihilation channel parameterized by $\kappa$; with these assumptions the only dependence of the annihilation rate on $m_\phi$ and $\delta$ comes through the low-velocity Sommerfeld enhancement, so except near the resonance centers, the relic density is largely fixed by $\alpha_D$ and $m_\chi$. Since these resonance regions are generally already ruled out by constraints from the CMB, we simply hold $\alpha_D$ fixed for these scans, and confirm that an acceptable relic density is obtained in all the regions that are not ruled out. For the minimal model, in contrast, the early-universe cross section for the new annihilation channel $\chi_i \chi_i \rightarrow h_D h_D$ scales as $\delta^2/m_\phi^4$, so the coupling $\alpha_D$ must be reduced at small $m_\phi$ and/or large $\delta$ to obtain the correct relic density. 

A quick study of these plots shows that boosts larger than 200 populate a large region of the allowed parameter space, with boosts even larger than 300. Much larger boosts are still possible to achieve while maintaining agreement with the relic density constraint ($S\gsim 500$) but are strongly disfavored from CMB constraints (note also that not every point with a high BF will provide good fits to the CR data; in particular, the $\rho$ resonance leads to a pronounced dip in the leptonic fraction for $m_\phi \sim 0.8$ GeV that can make the spectrum too soft to provide a good fit). For low $m_\phi$ or large $\delta$, the $\chi \chi \rightarrow h h$ annihilation channel in the minimal model naturally reduces the annihilation rate both in the local halo and during the epoch of last scattering; increasing $\kappa$ has the same effect, but independent of $m_\phi$ and $\delta$. Either opens up new regions of parameter space at low mediator masses, which would be ruled out by the CMB constraints in the minimal model with $\delta=0$.

\subsection{Benchmark points}\label{sec:benchmarks}

We now provide specific examples of benchmark points that satisfy the relic density and CMB constraints.  Table \ref{tab:benchmarks} lists the particle physics parameters, present-day and low-velocity boost factors, and the CMB limit on the boost factor for these points. The boost factors are computed using the semi-analytic approximate formula for the Sommerfeld enhancement, but numerical checks indicate that the approximation is accurate to within $\sim 5\%$ for $v \gtrsim 150$ km/s, and to within $\sim 10\%$ for $v \rightarrow 0$. In both cases, the approximation tends to overestimate the enhancement, so the overall effect is to slightly weaken the CMB constraints relative to the present-day boost factor.

Two of our benchmarks have mass splittings well below $2 m_e$, and therefore potentially long lifetimes. We can estimate the depletion of the excited state due to DM-DM downscattering using the prescription for the scattering rate described in \S \ref{subsec:scattering}, although it should be noted that unlike the relic density calculation, the uncertainties in our prescription for the downscattering cross section induce correspondingly large uncertainties in the relic excited fraction. For the 1.68 TeV and 1 TeV benchmarks, the estimated relic excitation fractions (after decoupling of the DM-DM downscattering, but before decay) are respectively $\sim 5 \times 10^{-3}$ and $2 \times 10^{-3}$.

In Table \ref{tab:benchmarksGALPROP} we provide the \texttt{GALPROP} parameters for the electron and proton injection spectra needed to reproduce the background $\epm$ spectra for each of the benchmark points.  As discussed in \S \ref{sec:BGspectra}, the injection spectrum for electrons is a power law in energy with two breaks and so can be described by six parameters $(n_e, \gamma_{e1}, E_{e1}, \gamma_{e2}, E_{e2}, \gamma_{e3})$, an overall normalization $n_e$, the three power law indices, $\gamma_{e1}$, $\gamma_{e2}$, and $\gamma_{e3}$, and the energies at which the breaks occur $E_{e1}$ and $E_{e2}$.  The low energy break occurs around a few GeV.  Since our PAMELA and \emph{Fermi} fits do not extend to energies below 10 GeV, the electron spectrum at these energies is irrelevant for our analysis.  Therefore, the parameters $E_{e1}$ and $\gamma_{e1}$ are unnecessary for our fits, but we include them in Table \ref{tab:benchmarksGALPROP} for completeness.  The proton injection spectrum is a broken power law in energy described by the four parameters $(n_p, \gamma_{p1}, E_p, \gamma_{p2})$.

\begin{table*}[t]\footnotesize
\begin{tabular}{|c|c|c|c|c|c|c|c|c|}
\hline
Benchmark & Annihilation          & $m_\phi$  & $m_\chi$  & $\alpha_D$ & $\delta$ & Local & Saturated  & CMB  \\
number & channel &(MeV) & (TeV) & & (MeV) & BF & BF & limit \\
\hline
1 & 1:1:2 $e^\pm:\mu^\pm:\pi^\pm$ & 900      &  1.68       & 0.04067    & 0.15          & 300      & 530          & 600 \\
2 & 1:1:2 $e^\pm:\mu^\pm:\pi^\pm$ & 900      &  1.52       & 0.03725    & 1.34            & 260      & 360          & 545 \\
3 & 1:1:1 $e^\pm:\mu^\pm:\pi^\pm$ & 580      &  1.55       & 0.03523    & 1.49          & 250      & 437          & 490 \\
4 & 1:1:1 $e^\pm:\mu^\pm:\pi^\pm$ & 580      &  1.20       & 0.03054    & 1.00          & 244      & 374          & 379 \\
5 & 1:1 $e^\pm:\mu^\pm$                & 350       &  1.33       & 0.02643    & 1.10            & 156      & 339          & 340 \\
6 & $e^\pm$ only                             & 200       &  1.00       & 0.01622    & 0.70          & 67        & 171          & 171 \\
\hline
\end{tabular}
\caption{Particle physics parameters, present day boost factors, and boost factors in the low-velocity limit for benchmark points. The boost factor (BF) is defined as $\langle \sigma v \rangle / 3 \times 10^{-26}\rm cm^3 \;s^{-1}$.}
\label{tab:benchmarks}
\end{table*}
\begin{table*}[t]\footnotesize
\begin{tabular}{|c|c|c|c|c|c|c|c|c|c|c|}
\hline
Benchmark &  $n_e \times10^{-10}$ & $\gamma_{e1}$ &  $E_{e1}$ & $\gamma_{e2}$ & $E_{e2}$ & $\gamma_{e3}$ & $n_p \times10^{-9}$ & $\gamma_{p1}$ & $E_{p1}$ & $\gamma_{p2}$ \\
number & @ 34.5 GeV & & (GeV) & & (GeV) & & @ 100 GeV & & (GeV) & \\
\hline
1 & 3.12763 & -1.60 & 4.0 & -2.45 & 2200 & -3.3 & 5.66361 & -1.98 & 9.0 & -2.11 \\
2 & 3.12763 & -1.60 & 4.0 & -2.45 & 2200 & -3.3 & 5.66361 & -1.98 & 9.0 & -2.11 \\
3 & 3.12763 & -1.60 & 4.0 & -2.45 & 2200 & -3.3 & 5.66361 & -1.98 & 9.0 & -2.11 \\
4 & 3.06444 & -1.60  & 4.0 & -2.50 & 2200 & -3.3 & 5.20440 & -1.98 & 9.0 & -2.11 \\
5 & 3.09604 &  -1.60 & 4.0 & -2.45 & 2200 & -3.3 & 5.74014 & -1.98 & 9.0 & -2.11 \\
6 & 3.10299 &  -1.60 & 4.0 & -2.45 & 2200 & -3.3 & 5.74014 & -1.98 & 9.0 & -2.11 \\
\hline
\end{tabular}
\caption{\texttt{GALPROP} parameters describing the electron and proton injection spectra for the benchmark points.  For all cases the diffusion parameters used are the following: $D_0=4.00$ (multiplied by $10^{28} \cm^2 \s^{-1}$ to obtain $D(E)$ at $E=4$ GeV), $\alpha=0.50$, $h_d=4.0\kpc$.  Normalizations $n_e$ and $n_p$ are in units of $\rm cm^{-2}\; s^{-1}\; sr^{-1}\; MeV^{-1}$.  Because we do not fit the PAMELA or \emph{Fermi} data below 10 GeV, the values of $E_{e1}$ and $\gamma_{e1}$ are unnecessary for our fits.  However, we list our default values here.}
\label{tab:benchmarksGALPROP}
\end{table*}

\begin{figure}[h]
\centering
\begin{tabular}{ccc}
\includegraphics[width=0.30\textwidth]{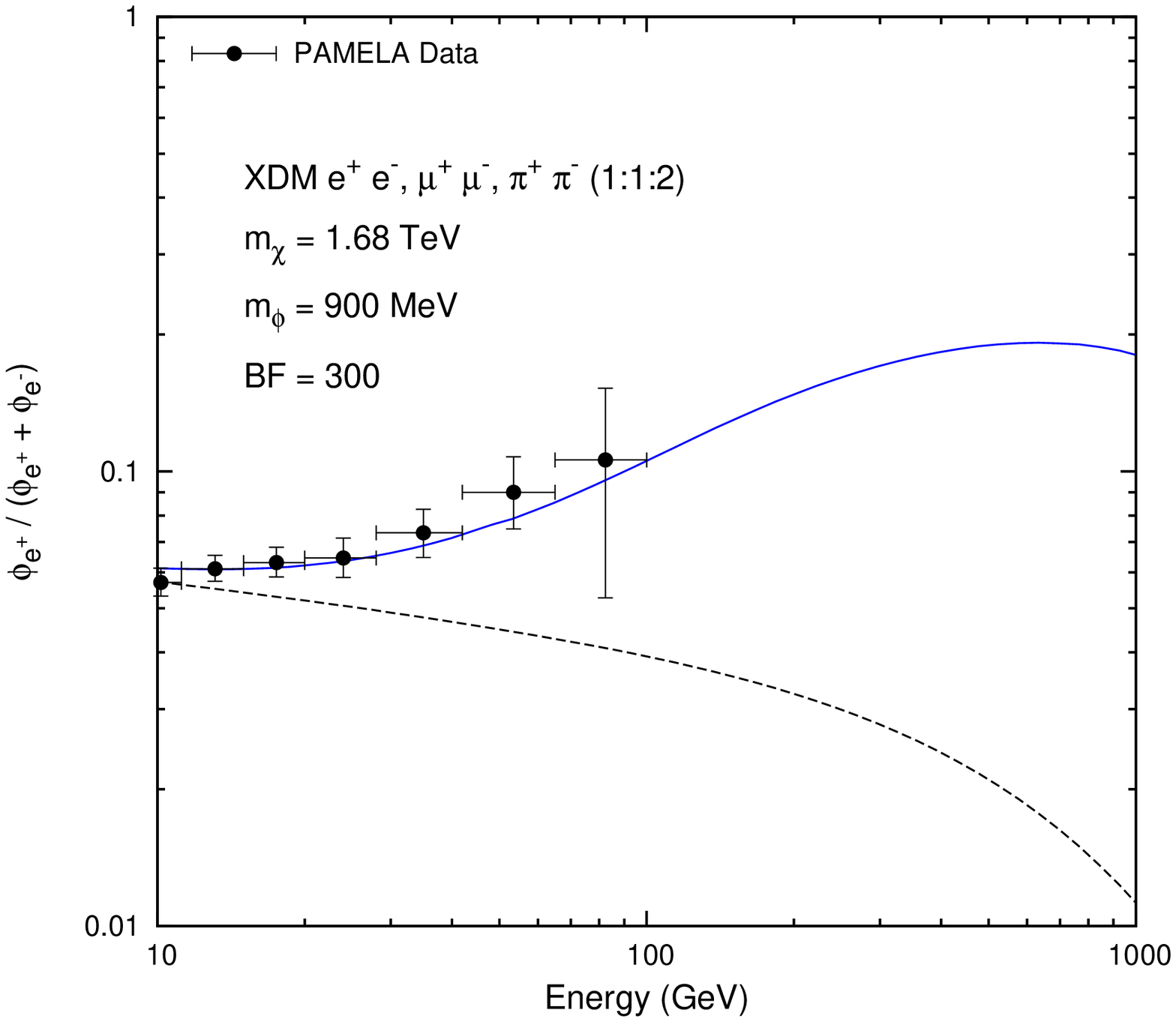} &
\includegraphics[width=0.30\textwidth]{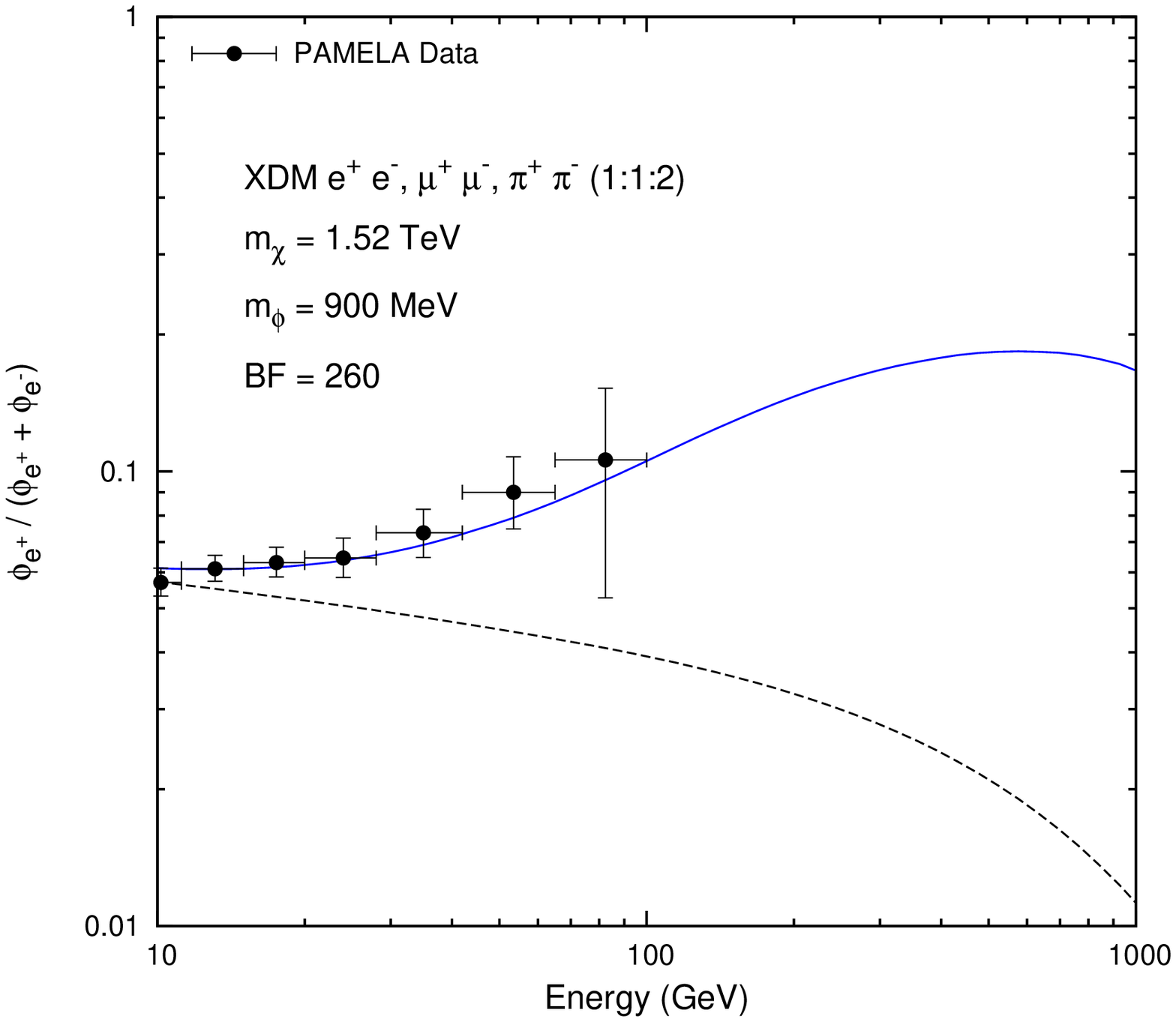} &
\includegraphics[width=0.30\textwidth]{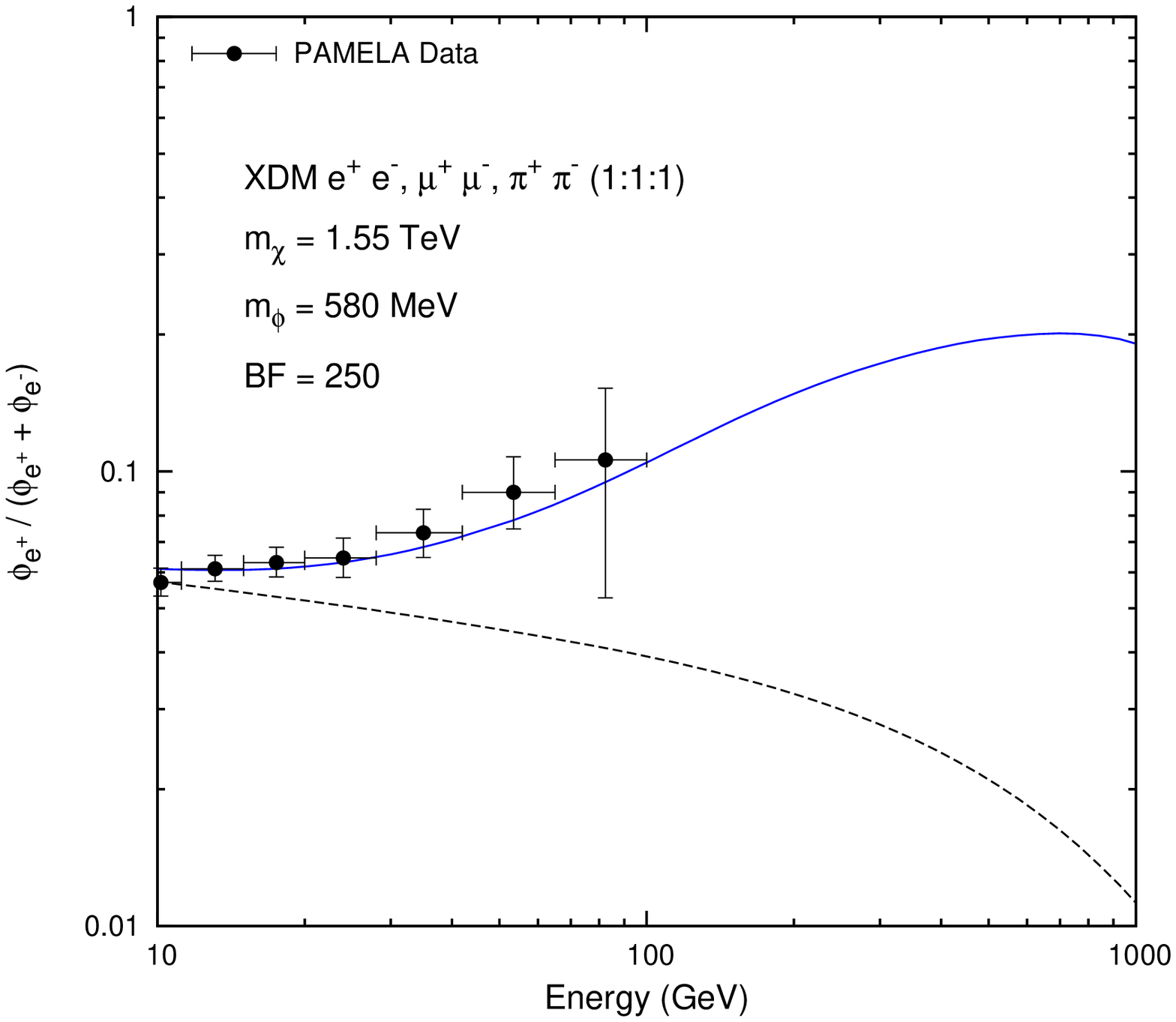} \\
\vspace{2pt}
\includegraphics[width=0.30\textwidth]{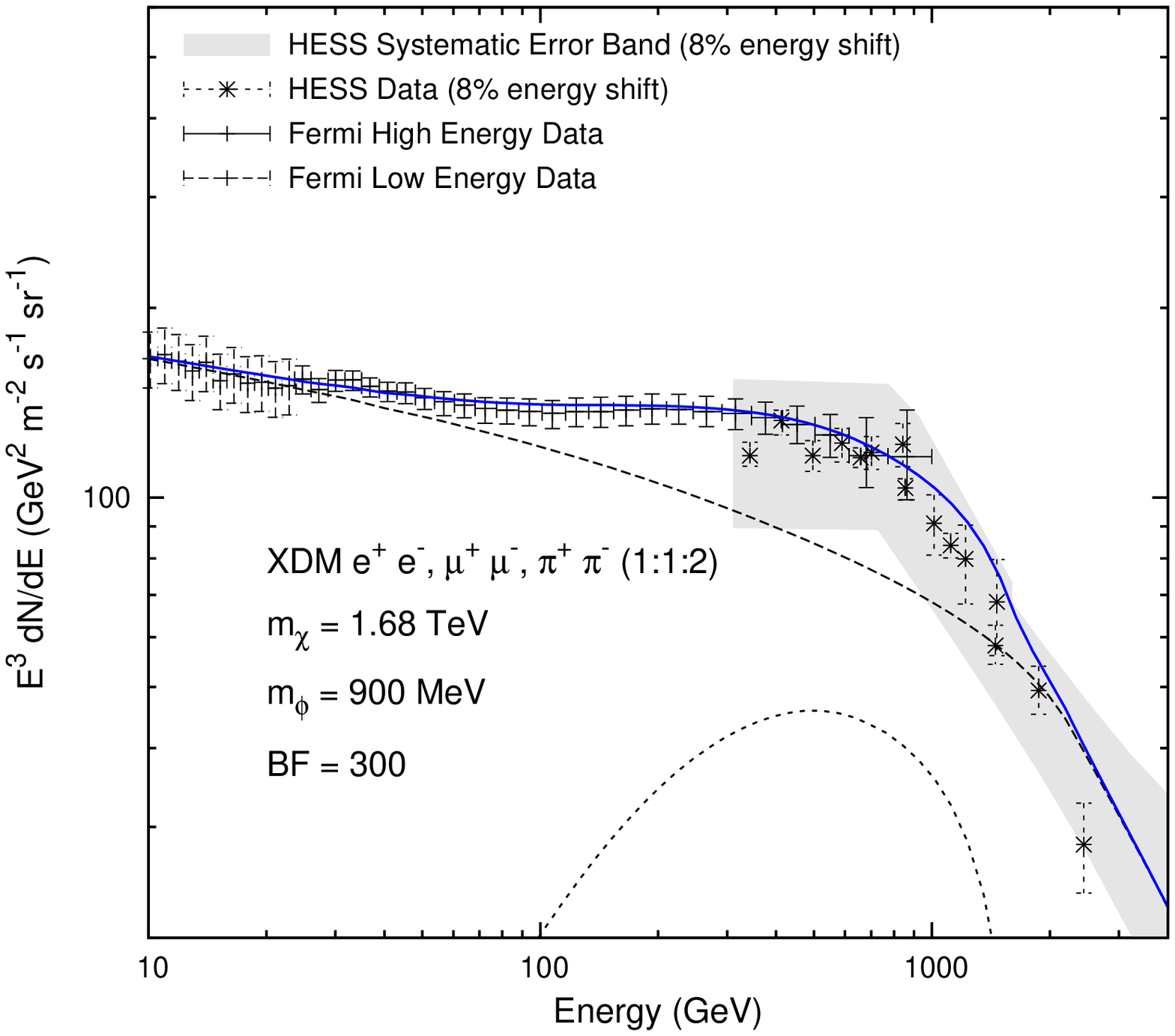} &
\includegraphics[width=0.30\textwidth]{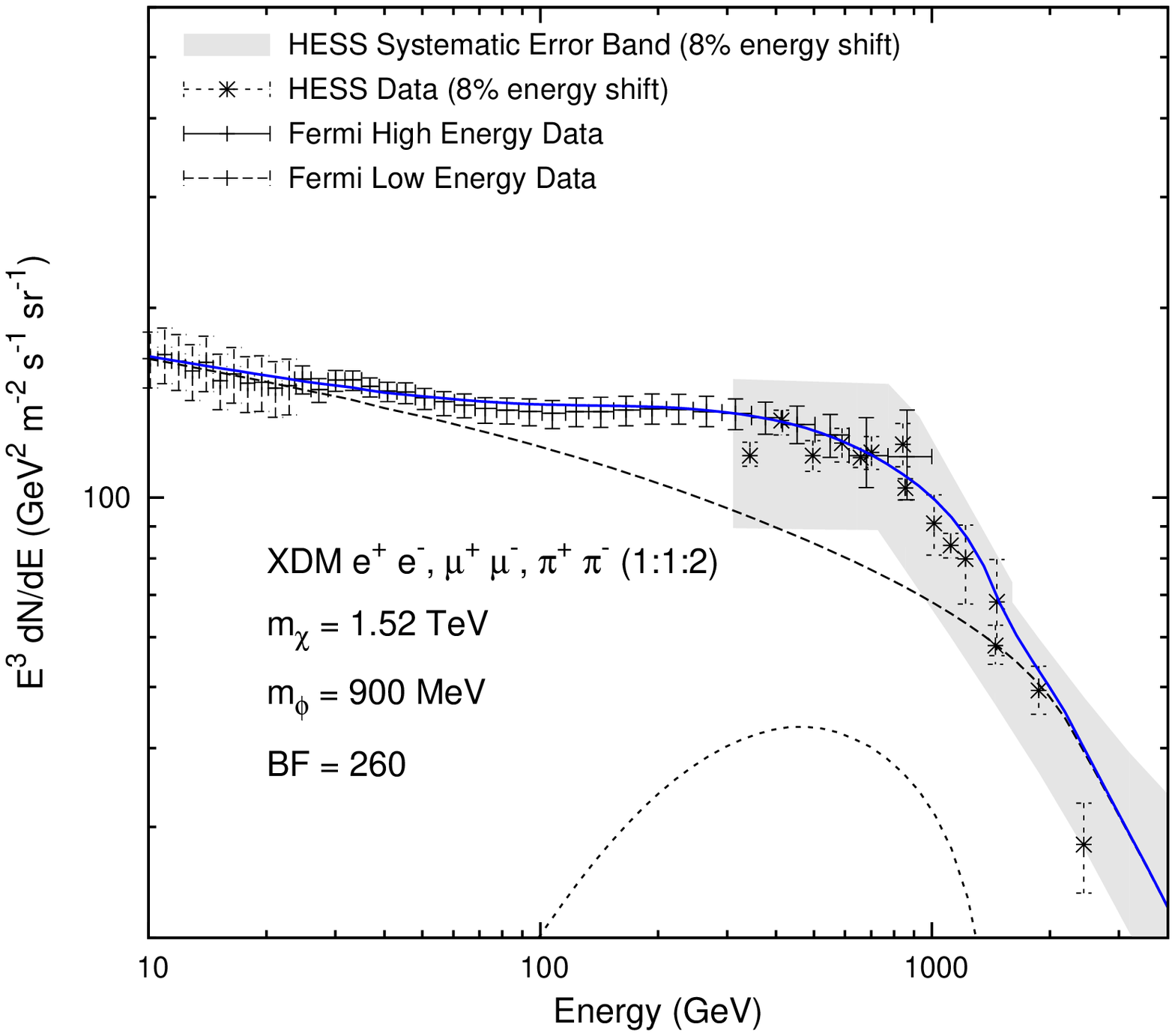} &
\includegraphics[width=0.30\textwidth]{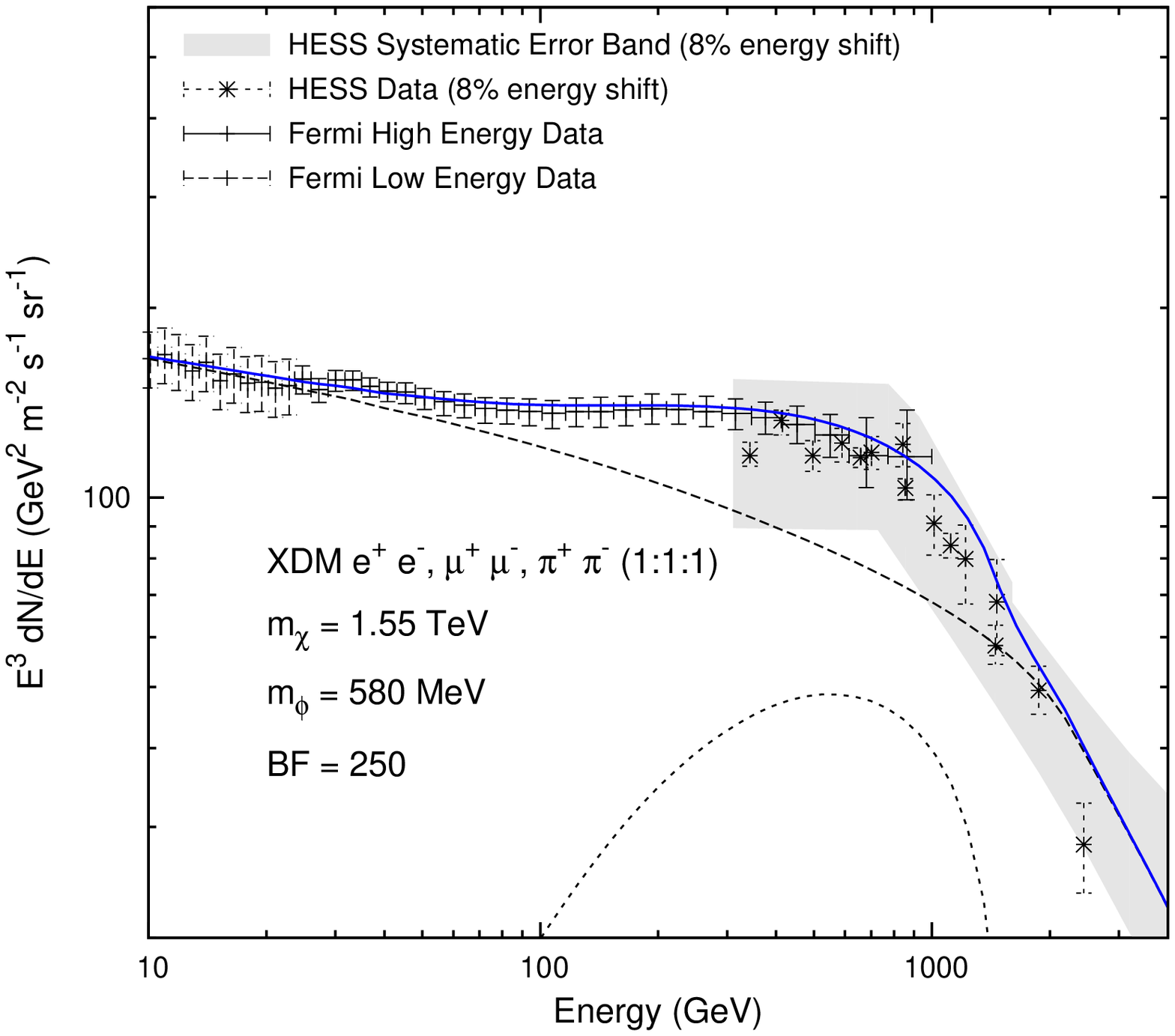} \\
\vspace{2pt}
\includegraphics[width=0.30\textwidth]{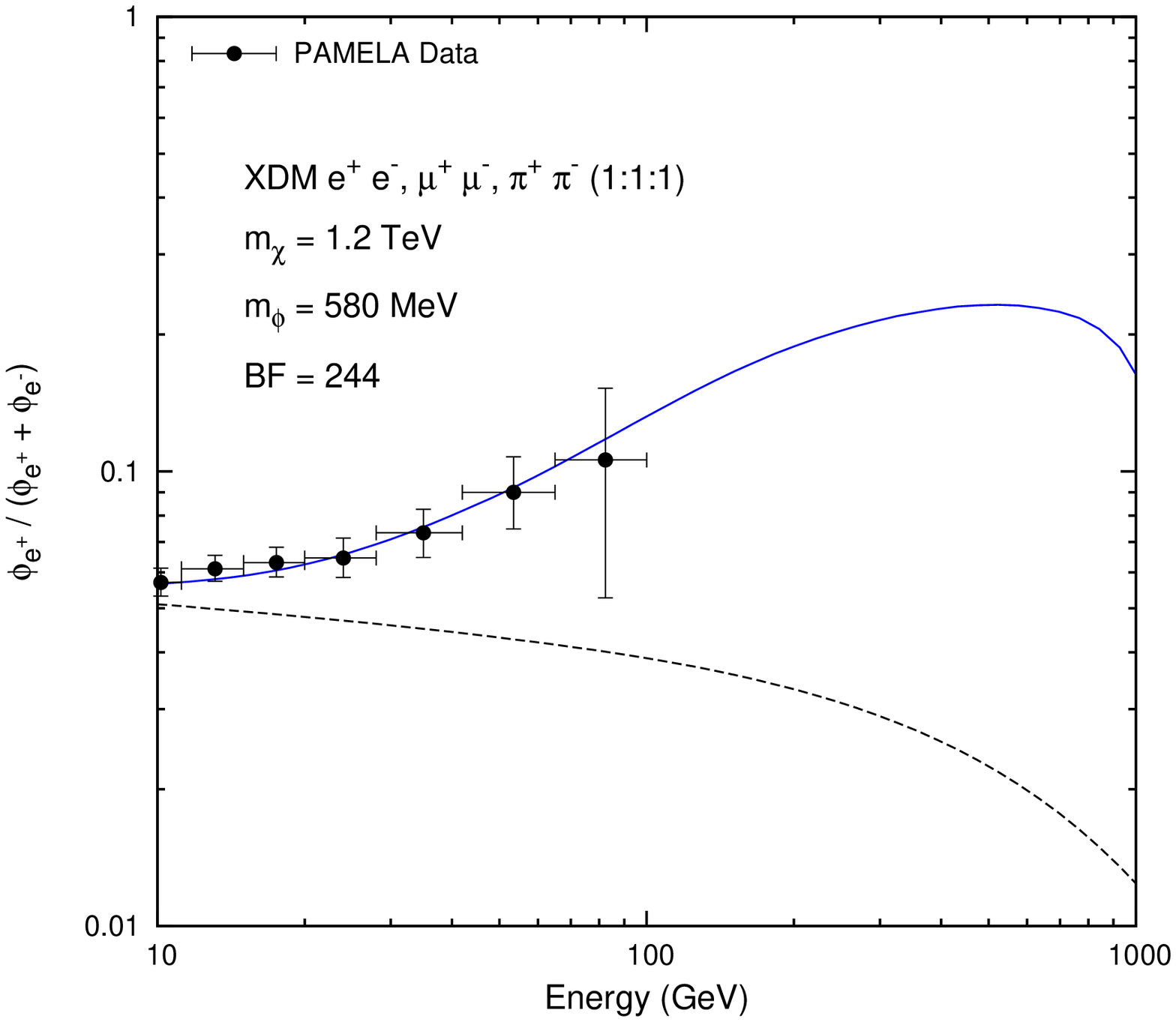} &
\includegraphics[width=0.30\textwidth]{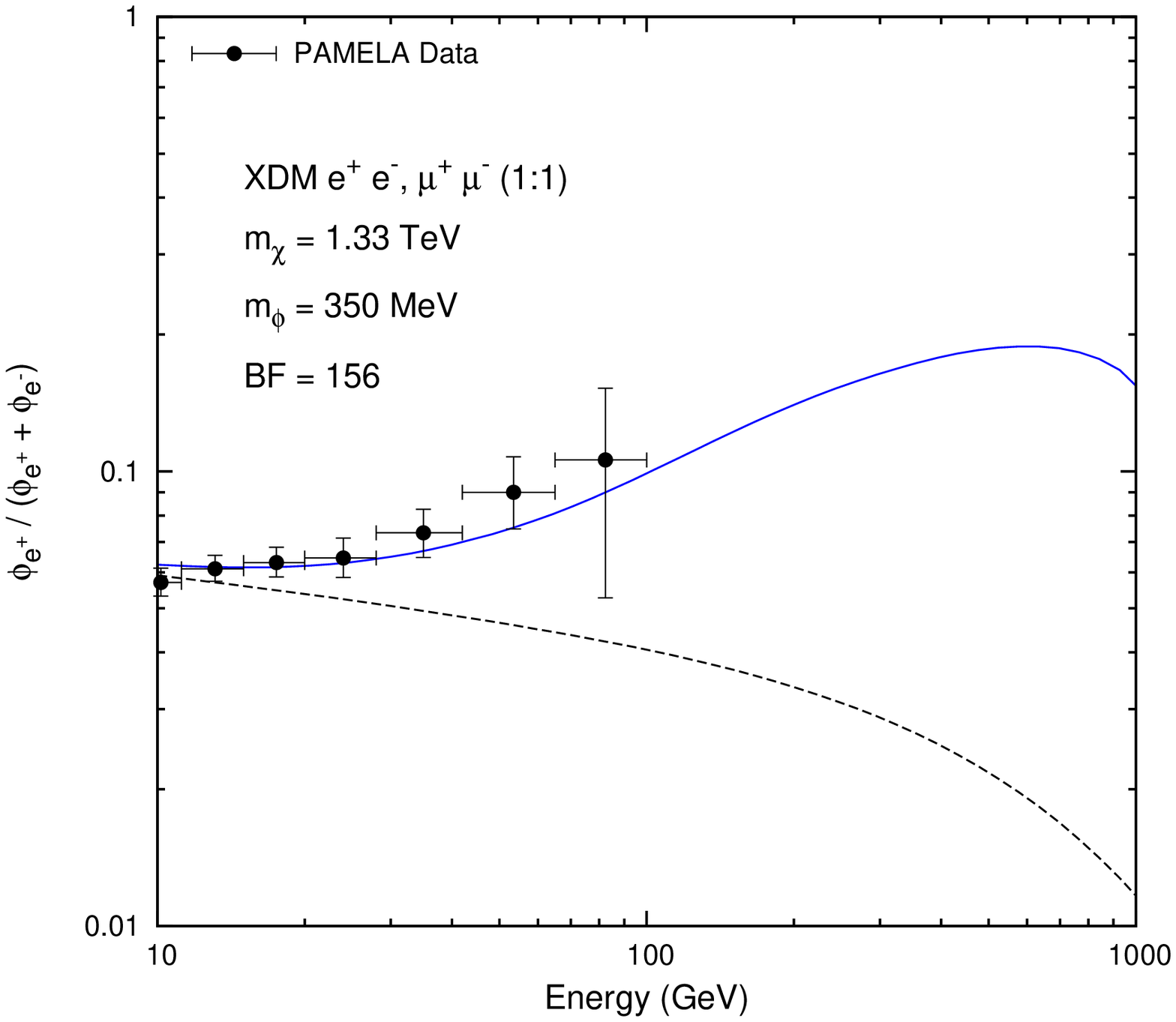} &
\includegraphics[width=0.30\textwidth]{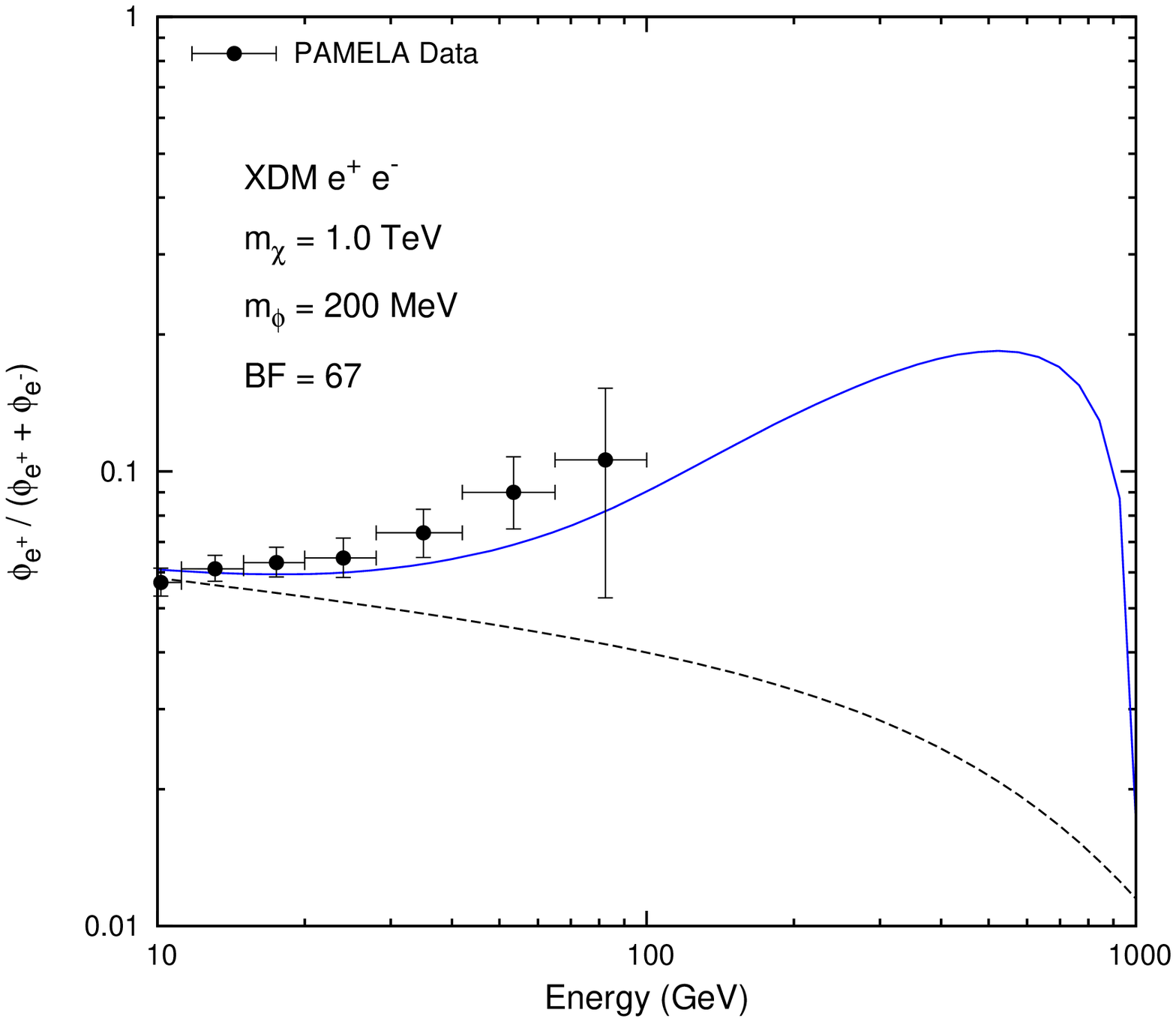} \\
\vspace{2pt}
\includegraphics[width=0.30\textwidth]{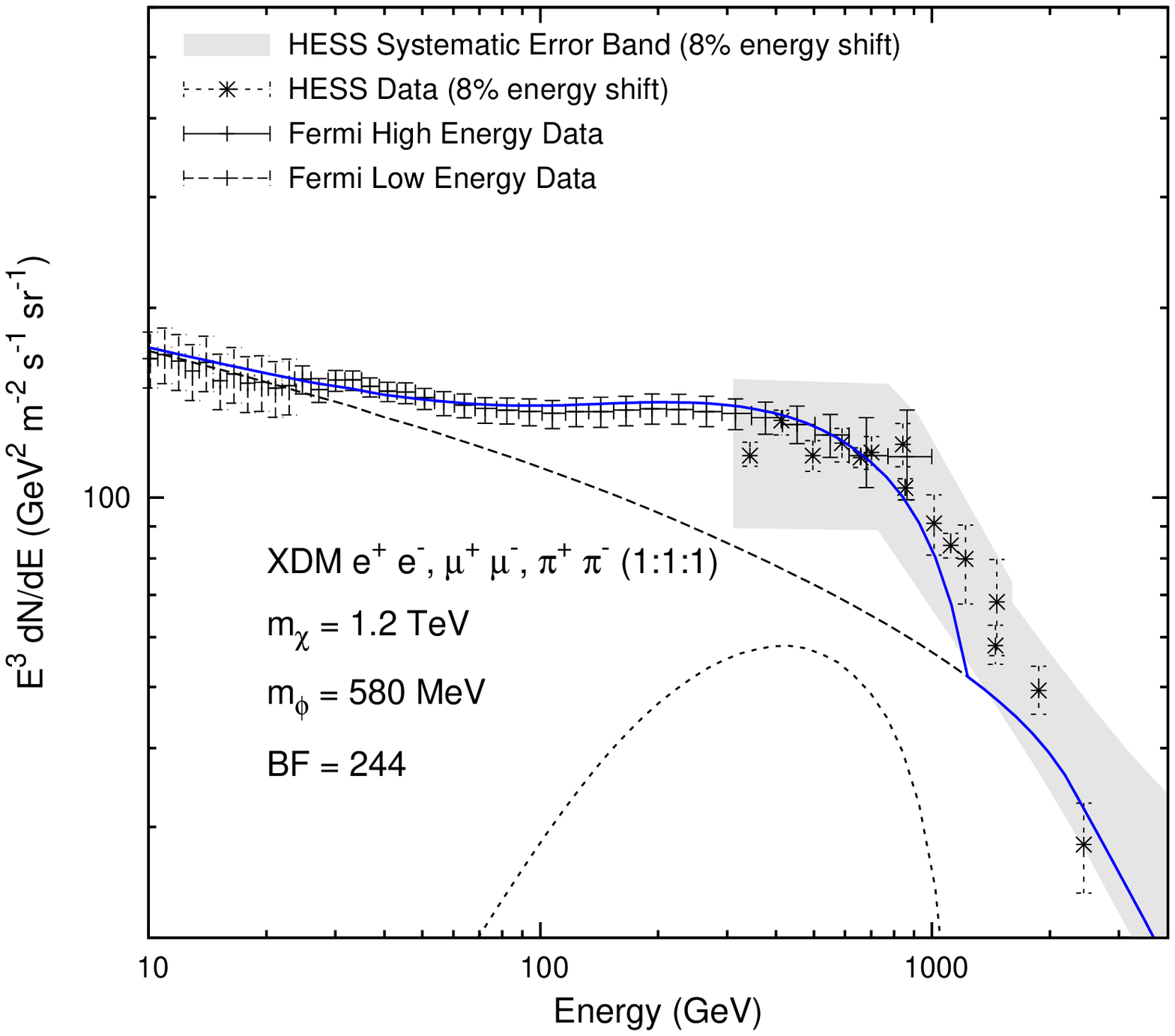} &
\includegraphics[width=0.30\textwidth]{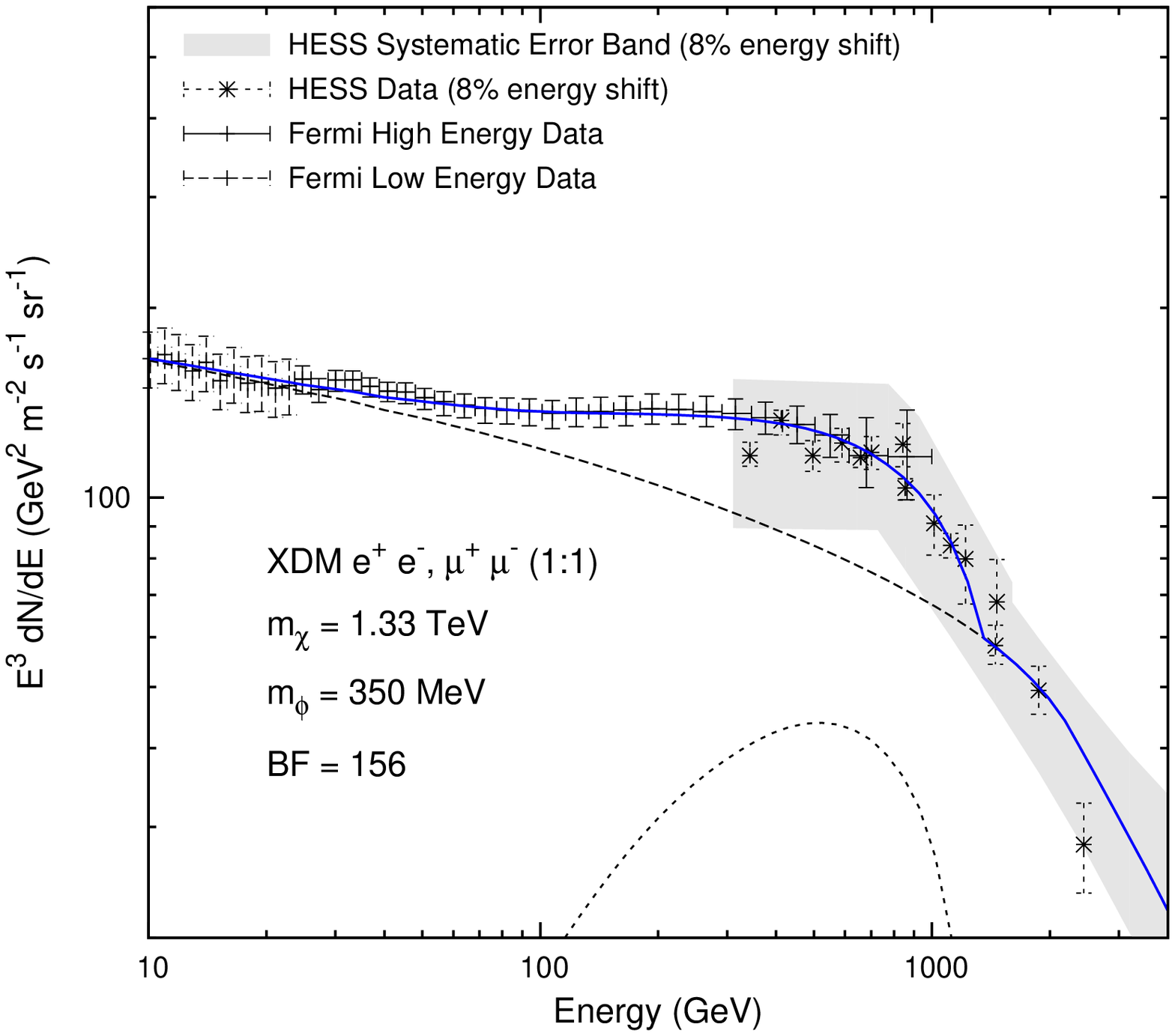} &
\includegraphics[width=0.30\textwidth]{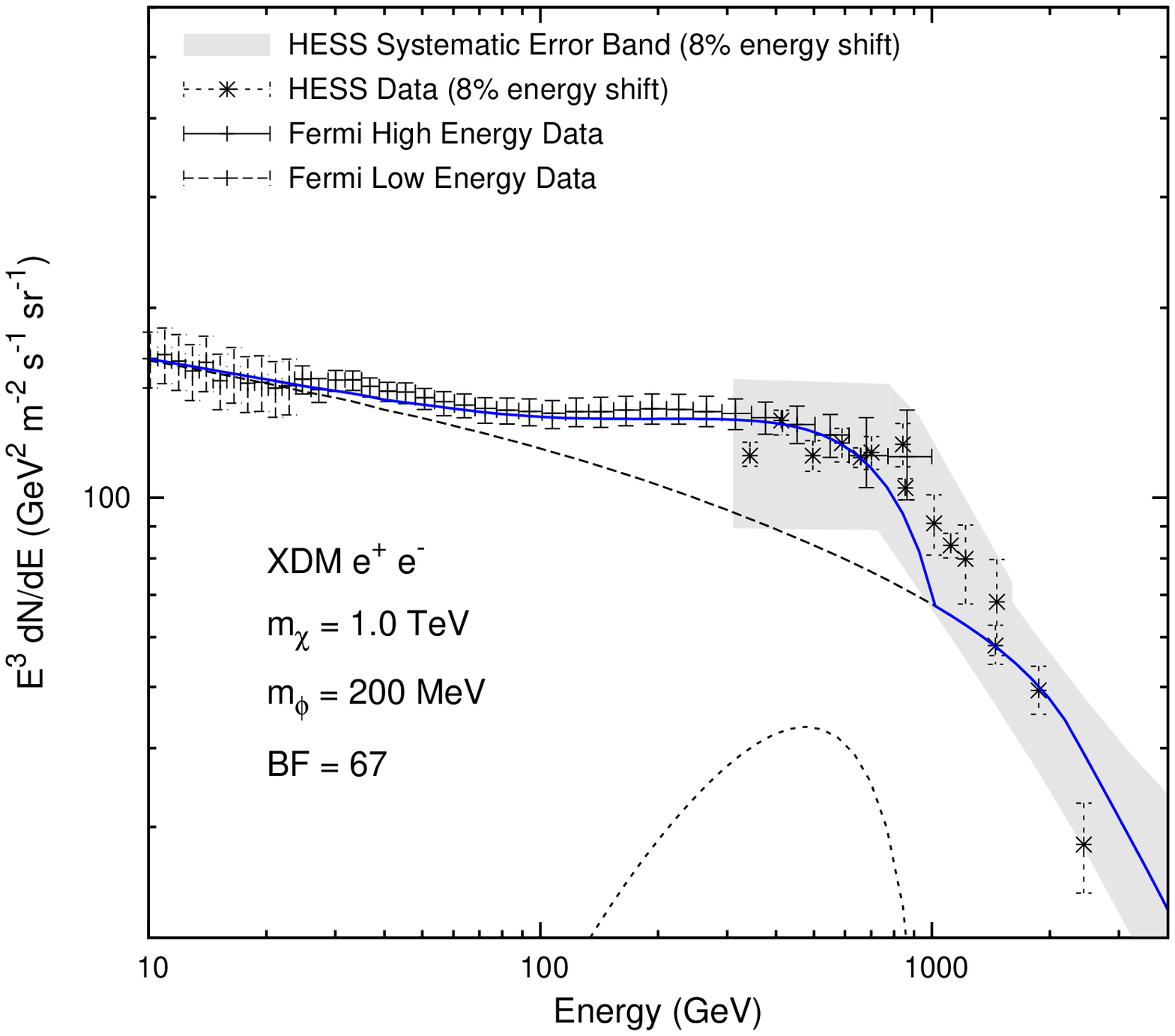} \\
\end{tabular}
\caption{\footnotesize Benchmark models fitting the PAMELA (\emph{first and third rows}) and \emph{Fermi} (\emph{second and fourth rows}) cosmic-ray excesses, obtained using the \texttt{GALPROP} program.}
\label{fig:benchmarks}
\end{figure}

\subsection{Comparisons with Previous Results}
While we find ample regions of parameter space that provide agreement with the PAMELA and \emph{Fermi} results, previous studies \cite{Feng:2009hw} have been more negative. In particular, \cite{Feng:2009hw} finds a maximum local ``boost factor'' (BF) of $\sim 120$ from Sommerfeld-enhanced annihilation for $\sim 2$ TeV DM, and a BF of 90 for 1 TeV DM, compared to a best fit to the data for $\chi \chi \rightarrow \phi \phi$, with $\phi \rightarrow \mu^+\mu^-$ (taken from \cite{Bergstrom:2009fa,Meade:2009iu}) of 2.35 TeV DM with a boost factor of 1500. We should emphasize that our results are completely consistent with theirs, with the different conclusions arising from our consideration of a more general parameter space. Specifically,

\begin{itemize}
\item{A decay mode of  $\phi\rightarrow \mu^+\mu^-$ was assumed by \cite{Feng:2009hw}, which is natural for models where $\phi$ is a scalar, but does not occur in models where $\phi$ is a vector, and the force arises from a conventional gauge group. In these cases, $\phi$ couples to charge, and there is always a sizeable hard $\phi\rightarrow e^+e^-$ component, unless $\phi$ is very degenerate with the $\rho$ meson. The presence of an electron component hardens the final $e^+ e^-$ spectrum, increases the power in $e^+ e^-$ as opposed to neutrinos (by a factor of up to $\sim 3$, depending on the branching ratio), and lowers the preferred mass scale for the DM to around 1 TeV, since the electron component dominates the high-energy cutoff behavior which sets the mass scale, both of which lower the needed boost.}
\item{For vector mediators, it is both natural and almost an experimental necessity (given constraints from direct detection) to consider the case where $\chi_1$ and $\chi_2$ are non-degenerate. Such a scenario generally produces larger Sommerfeld enhancements than in the degenerate case.}
\item{Older works on the local DM density \cite{Gates:1995dw} found a central value of $\rho_0 = 0.3 \gev/{\rm cm^3}$, a number which has become a standard. As we noted above, more recent studies, however, find values of $0.39$ \cite{Catena:2009mf}, $0.43$  \cite{Salucci:2010qr} and $0.46$ GeV/cm$^3$ \cite{Pato:2010yq}. Taking the current best estimates of the local density then leads to a factor of $\gsim 2$ reduction in the required boost factor.}
\end{itemize}

The combination of these effects suggests that the natural boost required for the majority of the parameter space is $\mathcal{O}$(100-300), or a factor of $\mathcal{O}(5)$ lower than the models considered in \cite{Feng:2009hw}, or an order of magnitude (relative to \cite{Bergstrom:2009fa,Meade:2009iu}) when combined with the best current estimates for the local relic density.  Our results are consistent with the conclusions of \cite{Feng:2009hw} that $\chi \chi \rightarrow \phi \phi$, $\phi \rightarrow \mu^+\mu^-$ is tightly constrained, but the general parameter space (in which some $\phi \rightarrow \epm$ is present) is far less so. This conclusion need not invoke artificially high local substructure, but only a more accurate estimate of the preferred boost factor and DM mass range for the specific class of scenarios we consider, where the DM annihilates through a dark gauge boson that kinetically mixes with SM hypercharge. 

\section{Conclusions}
Data from several experiments have pointed to the presence of a new, primary source of cosmic ray electrons and positrons. Dark matter is a long-standing candidate for such a source, but has difficulty achieving the high rates, hard spectrum and dearth of correlated anti-protons that are required by the data. Models of dark matter with a new, light boson $\phi$ can qualitatively explain such phenomena through annihilations $\chi \chi \rightarrow \phi \phi$, followed by $\phi \rightarrow e^+e^-, \mu^+\mu^-, \pi^+\pi^-$. 

We have seen that this agreement extends to fully quantitative connections as well. Significant regions of parameter space exist for general models where the present-day boost is large enough to explain the observed cosmic-ray signals, while yielding the appropriate relic density and evading CMB constraints. We have additionally specified benchmark points which are representative of a sizeable fraction of the parameter space, in order to give a first set of parameters to be searched for in terrestrial experiments.

In general, the allowed regions of parameter space have a non-trivial $\phi\rightarrow e^+e^-$ branching ratio -- models that go exclusively into $\mu^+\mu^-$ have difficulty in achieving the necessary boost while being consistent with other constraints, although a proper treatment of capture into bound states can alleviate this tension for DM masses above 2 TeV. At the same time, models that go {\em exclusively} to $e^+e^-$ are more severely constrained by limits from the CMB, as mediators lighter than $2 m_\mu$ often yield too {\em large} a signal in the epoch of recombination. This suggests a preferred mass range of $2 m_\mu < m_\phi \lsim 1 {\rm GeV}$.

Fortunately, this range of parameters is very testable. Low-energy experiments \cite{Batell:2009yf,Essig:2009nc,Bjorken:2009mm,Essig:2010xa,AmelinoCamelia:2010me}, such as APEX should be able to study a wide range of parameters, while LHC and Tevatron searches \cite{Baumgart:2009tn,Abazov:2010uc} offer complementary reach. Finally, for all of these models, the Planck satellite should see a signal in its polarization spectrum.
In light of this, these models remain a viable, testable and thus exciting scenario for dark matter. 

As a resource for interested readers, a web application for the calculations given in this work can be found at \texttt{http://astrometry.fas.harvard.edu/mvogelsb/sommerfeld/}. It allows the user to compute the present-day and saturated boost factors, the relic density, and an estimate for the relic abundance of the excited state (prior to any decay), for our minimal model with any choice of parameters ($\alpha, m_\chi, m_\phi, \delta$), with the option to include WIMPonium formation in the $\delta=0$ case. It also automates the calculation of the semi-analytic approximation for the Sommerfeld enhancement, for a $U(1)_D$ model, with the initial-state particles in either the ground or excited state. Readers wishing to run large parameter scans or explore options not available in the web application should contact TS at \texttt{tslatyer@ias.edu} for access to the underlying code.

\acknowledgments

We acknowledge helpful conversations with Jolyon Bloomfield, Rouven Essig, Jonathan Feng, Manoj Kaplinghat, Mariangela Lisanti, Maxim Pospelov, Josh Ruderman, Philip Schuster, Natalia Toro, and Tomer Volansky. DF and TS received partial support from NASA Theory grant NNX10AD85G.  TS was partially supported by a Sir Keith Murdoch Fellowship from the American Australian Association during the early phases of this work, and gratefully acknowledges the hospitality of the Aspen Center for Physics, and support from the Institute for Advanced Study.  In the later phases of this work, the research of TS was supported by DOE grant \#DE-FG02-90ER40542 and NSF grant \#AST-0807444. NW is supported by DOE OJI grant \#DE-FG02-06ER41417 and NSF grant \#0947827, as well as support from the Amborse Monell Foundation.

\appendix

\section{Capture into WIMPonium and Heavy DM Scenarios}
\label{sec:wimponium}

The usual Sommerfeld enhancement calculation neglects diagrams describing the capture of two DM particles into a bound state -- referred to as WIMPonium, in analogy to positronium -- accompanied by the radiation of a dark gauge boson. This process is kinematically allowed if,
\begin{equation} m_\phi < \alpha_D^2 m_\chi / 4 + 2 \left( \sqrt{m_\chi^2 + |\vec{p}|^2} - m_\chi \right), \end{equation}
which at low velocities, $T \lesssim m_\phi$, reduces to $m_\phi < \alpha_D^2 m_\chi /4$ as stated in e.g. \cite{Pospelov:2008jd}.

Above the symmetry breaking scale, the $\phi$ is massless and this process is exactly analogous to positronium formation with the replacement $m_e \rightarrow m_\chi$, $\alpha_\mathrm{EW} \rightarrow \alpha_D$. The nonrelativistic cross section (valid in the regime $m_\phi \lesssim T \lesssim m_\chi$) is given approximately by \cite{0953-4075-29-10-021},
\begin{equation} \sigma \approx \left\{ \begin{array}{cc} \frac{2^9 }{3}  \left(\frac{\omega}{\alpha_D^2 m_\chi/4}\right) \left(\frac{\pi \alpha_D^2}{m_\chi^2 v} \right) \left( \frac{\pi \alpha_D/v}{1 - e^{-\pi \alpha_D /v}} \right) \left(\frac{(\alpha_D/2 v)^2}{1+(\alpha_D/2v)^2}\right)^3 & \\
\times e^{-4 \tan^{-1} (\alpha_D / 2 v)} \left( 1 + \frac{\omega^2 (1 - (\alpha_D / 2 v)^2)}{5 m_\chi^2 v^2} \right), & v \gtrsim \alpha_D/2, \\ \frac{2^9 e^{-4}}{3} \left(\frac{\pi \alpha_D^2}{m_\chi^2 v} \right) \left(\frac{\pi \alpha_D}{v}\right) \left(1 - \frac{8}{3} \left(\frac{v}{\alpha_D}\right)^2 - \frac{16}{5} \left(\frac{\omega}{\alpha_D m_\chi}\right)^2 \right), & v \ll \alpha_D/2, \end{array} \right. \end{equation}
where $\omega$ is the energy of the radiated $\phi$, $\omega \approx \alpha_D^2 m_\chi/4 + m_\chi v^2$, again assuming $v \ll 1$. This expression agrees with \cite{Pospelov:2008jd} in the low-velocity limit when $m_\phi$ is set to zero. The capture cross section experiences Sommerfeld enhancement just as the annihilation cross section does, but relative to the annihilation cross section is greatly suppressed at high velocities, with $\sigma v$ scaling as $(2 v/\alpha_D)^{-4}$ for $v/\alpha_D \gtrsim 1/2$. At low velocities, on the other hand, the capture cross section scales exactly as the Sommerfeld-enhanced direct annihilation. 

This behavior can be understood from the relative ranges of the various interactions: the length scale associated with the annihilation operator is 1/$m_\chi$ (determined by the mass of the $\chi$), whereas for the capture operator it is 1/$\alpha_D m_\chi$ (determined by the Bohr radius of the WIMPonium). The momentum of the incoming particles is $m_\chi v$: for nonrelativistic particles, this is never large enough to probe the $r \lesssim 1/m_\chi$ region relevant for annihilation, but it is large enough that for $v \gtrsim \alpha_D$, capture cannot be treated as a contact interaction. For $v \lesssim \alpha_D$, on the other hand, both capture and annihilation behave as contact interactions, and so in both cases the Sommerfeld enhancement depends only on the behavior of the wavefunction at the origin. At least in the elastic ($\delta=0$) case, the shape of the wavefunction at the origin is completely determined by the requirement that it be finite, with the longer-range physics only setting its amplitude: consequently, the enhancement from the long-range interaction must scale in the same way for any operators localized at the origin. At energies below $m_\phi$, therefore, we employ the expression for WIMPonium capture given by \cite{Pospelov:2008jd}, but with the replacement $\pi \alpha_D/v \rightarrow S$, where $S$ is the Sommerfeld enhancement due to the Yukawa potential:
\begin{equation} \langle \sigma v_\mathrm{rel} \rangle = \frac{2^{10} e^{-4}}{3} \left(\frac{\pi \alpha_D^2}{m_\chi^2} \right) \left(\frac{v_\phi (3 - v_\phi^2)}{2} \right) S, \quad v_\phi = \sqrt{1 - (4 m_\phi / \alpha^2 m_\chi)^2}. \end{equation}

One might ask whether it is justified to neglect WIMPonium capture as we have done in previous sections. If $m_\phi > \alpha_D^2 m_\chi/4$, i.e. the capture process is kinematically forbidden at low velocities, then $v \sim \alpha_D/2$ corresponds to $T \sim \alpha_D^2 m_\chi/4 < m_\phi$. Thus the strong $(2 v/\alpha_D)^{-4}$ suppression of the capture cross section holds for all temperatures above the symmetry breaking scale.

Furthermore, for $T \gtrsim m_\phi$, interactions with the bath of relativistic $\phi$'s can dissociate the positronium \cite{Pospelov:2008jd}, so the timescale for a WIMPonium bound state to undergo $\phi$-dissociation must be compared to the annihilation rate. Only at $T \lesssim m_\phi$ can the capture cross section be used as a proxy for the annihilation cross section, since any particles that form a bound state will eventually annihilate.

The lifetime of positronium is $\tau \approx 1.2 \times 10^{-10}$ s for parapositronium ($1/4$ of the total positronium formed) and $\tau \approx 1.4 \times 10^{-7}$ s for orthopositronium ($3/4$), scaling as $1/m_e \alpha_\mathrm{EW}^5$ and $1/m_e \alpha_\mathrm{EW}^6$ respectively \footnote{The lifetime of orthoWIMPonium can be much shorter than expected from this scaling relation -- comparable to that of paraWIMPonium -- if the $\phi$ can decay into other dark-sector states, due to the availability of $s$-channel annihilation through an off-shell $\phi$, in analogy to true (ortho)muonium decay. However, in this analysis we will -- as previously discussed for the $\delta=0$ case -- neglect interactions between the $\phi$ and other dark-sector states, such as the dark Higgs.}. The lifetimes for the analogous WIMP states are shorter by a factor of $(m_e/m_\chi) (\alpha_\mathrm{EW}/\alpha_D)^{5,6}$. The photodissociation cross section for positronium is given by $\sigma \approx 2.5 \times 10^{-17}$ cm$^2$ at threshold, falling roughly as energy$^{-3}$ above threshold, and scaling as $\alpha_\mathrm{EW} a_0^2 \sim 1/\alpha_\mathrm{EW} m_e^2$ \cite{1989ApJ...344..232G}; for an initial estimate, we will just use the threshold value, since at earlier times the increased $\phi$ density will cancel out the reduced cross section. The $\phi$ density $n_\phi$ is obtained from Equation \ref{eq:yeq}, and can be written $n_\phi \approx 10^{40} (m_\phi/1\mathrm{GeV})^3$ cm$^{-3}$ when $T \sim m_\phi$.

Photodissociation and decay are equally likely at $T \sim m_\phi$ when $n_\phi \sigma v \approx 1/\tau$: this relation is equivalent to the conditions,
\begin{equation} \frac{m_\phi}{\alpha_D^2 m_\chi/4} \approx 1, \quad \mathrm{paraWIMPonium}, \end{equation}
\begin{equation} \frac{m_\phi}{\alpha_D^2 m_\chi/4}  \approx 0.1  \left(\frac{\alpha_\mathrm{EW}}{\alpha_D}\right)^{-1/3}, \quad \mathrm{orthoWIMPonium}. \end{equation}
Thus photodissociation is efficient at suppressing paraWIMPonium decay for $m_\phi \gg \alpha_D^2 m_\chi/4$ (the ratio of the rates near threshold scales as $m_\phi^3$), and always dominates orthoWIMPonium decay unless $m_\phi \ll \alpha_D^2 m_\chi/4$. 

The combination of this effect with the suppressed high-velocity cross section justifies our neglect of WIMPonium capture for $m_\phi > \alpha_D^2 m_\chi/4$, which holds true for all our benchmark points and the bulk of the parameter space we consider (with DM masses in the $1-2$ TeV range). More generally, for DM masses in this range, regions where low-velocity capture is allowed are mostly already ruled out by constraints from the CMB, as shown in Figure \ref{fig:paramscan}. 

However, $\alpha_D$ scales very roughly as $m_\chi$ in models with the correct relic density, so the threshold mediator mass $m_\mathrm{thres} = \alpha_D^2 m_\chi/4 \propto m_\chi^3$. For $m_\chi \gtrsim 2$ TeV, if there are no additional annihilation channels to reduce the required value of $\alpha_D$, $m_\mathrm{thres} \gtrsim 1$ GeV, and it becomes critical to take WIMPonium formation into account for mediator masses in the entire sub-GeV range.

For the purpose of studying WIMPonium formation, we restrict ourselves to the elastic case ($\delta = 0$), as previously neglecting interactions involving the dark Higgs. As we have argued above, we expect the low-velocity capture cross section to trace the Sommerfeld enhancement, just with a different prefactor (higher by a factor of $\sim 7.25$). If this assumption is valid, the boost factor in the low-velocity limit still scales as $\alpha_D m_\chi / m_\phi \propto m_\chi^2 / m_\phi$, so for $m_\phi$ close to the threshold value, $\mathrm{BF}_\mathrm{sat} \propto m_\chi^{-1}$, whereas the CMB constraint on this quantity scales as $m_\chi$. Thus for mediators with masses calibrated to the WIMPonium threshold, the CMB limits become much \emph{less} constraining at higher DM masses, and for $m_\chi \gtrsim 2$ TeV, we expect to find CMB-allowed regions of parameter space where WIMPonium formation cannot be neglected.

\begin{figure}[t]
\centering
\includegraphics[width=0.5\textwidth]{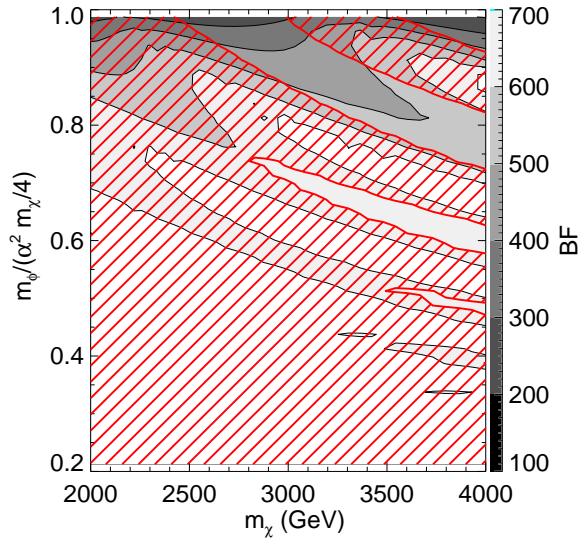}
\caption{\footnotesize The present-day boost factor for the case of a single DM state with $m_\phi$ light enough to permit radiative capture into WIMPonium, as a function of the DM mass $m_\chi$, and the mediator mass $m_\phi$ normalized to the WIMPonium binding energy. $\alpha_D$ is tuned to obtain the correct relic density. Red-hatched regions are ruled out at $95\%$ confidence by constraints from WMAP5, taking the energy deposition fraction $f=0.2$ to be conservative.}
\label{fig:wimponium}
\end{figure}

Multi-TeV candidates to fit the cosmic-ray data must give rise to relatively soft $e^+ e^-$ spectra with small or zero branching ratios to electrons, to avoid overproducing electrons at energies extending up to the DM mass; these features are not easily obtained from the simple vector portal models we have considered here (however, they can be accommodated in related models, e.g. \cite{Fox:2008kb}). In turn, such models require significantly higher present-day boost factors, as discussed earlier in the context of \cite{Feng:2010zp} -- but conveniently, taking WIMPonium formation into account will inevitably give rise to just such larger boost factors. 

For illustration, we solve the Boltzmann equation including WIMPonium formation and $\phi$-dissociation for 2-4 TeV DM, assuming the WIMPonium capture rate to trace the Sommerfeld enhancement for $v \lesssim \alpha_D$ as discussed above, and taking the energy dependence of the $\phi$-dissociation cross section to be \cite{1989ApJ...344..232G}:
\begin{equation} \sigma(\epsilon) \propto \epsilon^{-4}\frac{e^{4(1 - \eta \cot^{-1} \eta)}}{1 - e^{-2\pi\eta}}, \quad \eta = (\epsilon -1)^{-1/2}, \quad \epsilon = E/E_\mathrm{thres}. \end{equation}
The results of this calculation are shown in Figure \ref{fig:wimponium}; we find that boost factors in the 600-700 range are possible if $m_\chi \gtrsim 2.8$ TeV, even for this elastic model (in comparison, the boost factor from Sommerfeld enhancement alone is $\sim 100$). Note that we neglect capture into excited states and transitions between the various bound states in the spectrum (for a discussion of such processes in the context of WIMPonium, see \cite{MarchRussell:2008tu}); these processes are not generally negligible, but are beyond the scope of this study. In much of the parameter space for few-TeV DM, in order to evade the CMB limits the $\phi$ mass must be quite close to the threshold value, and so capture into excited states will be kinematically forbidden at low velocities.

\section{The Origin of the Mass Splitting in a Simple Model}
\label{sec:makesplitting}

The high-energy Lagrangian in our ``minimal model'' takes the form,
\begin{align} \mathcal{L} & = i \bar{\Psi} \gamma^\mu \left( \partial_\mu + i g_D \phi_\mu \right) \Psi + \left(\partial^\mu + i g_D \phi^\mu \right) h_D \left(\partial_\mu - i g_D \phi_\mu \right) h_D^* - m_\chi \bar{\Psi} \Psi  \nonumber \\ & - \frac{y}{2 \Lambda} \left( \bar{\Psi^C} \Psi h_D^* h_D^* + \bar{\Psi} \Psi^C h_D h_D \right)  -\frac{1}{4} F^\mathrm{D}_{\mu \nu} F_D^{\mu \nu} - \frac{\epsilon}{2} F^\mathrm{EM}_{\mu \nu} F_D^{\mu\nu} + \mathcal{L}_\mathrm{SM}. \end{align}

We can use the high-energy limit to compute annihilation cross sections for the various processes that are approximately valid at all energies, since $m_\phi/m_\chi$ is small and in all cases there are leading order terms with no $m_\phi$ dependence. The annihilation of fermions and antifermions into gauge bosons and oppositely-charged scalars have both been computed elsewhere and we will not give details here. The annihilation induced by the operator $\bar{\Psi^C} \Psi h_D^* h_D^*$ and its conjugate is easy to compute: the matrix element for e.g. fermion-fermion annihilation is just $|\mathcal{M}| =| (y/\Lambda) u_1 \bar{v}_2|$ (where ``1'' and ``2'' label the ingoing particles). Averaging over initial spins yields,
\begin{equation} (1/4) \sum_{s_1,s_2} |\mathcal{M}|^2 = (1/4) (y/\Lambda)^2 \mathrm{Tr} \left(p_{1\mu} p_{2\nu} \gamma^\mu \gamma^\nu - m_1 m_2 \right) =(y/\Lambda)^2( p_1 \cdot p_2 - m_1 m_2), \end{equation}
and dividing by two to account for the two identical particles in the final state, we find the COM frame cross section,
\begin{equation} \sigma = \frac{2 \pi}{64 \pi^2 s} \frac{|\mathcal{M}|^2}{\sqrt{1 - 4 m_\chi^2/s}} =  \frac{1}{32 \pi s} \frac{s/2 - 2 m_\chi^2}{\sqrt{1 - 4 m_\chi^2/s}} \left(\frac{y}{\Lambda}\right)^2 = \frac{1}{64 \pi} \left(\frac{y}{\Lambda}\right)^2 \sqrt{1 - 4 m_\chi^2 /s}.\end{equation}

Now let us consider the symmetry breaking. We employ the notation of \cite{Dreiner:2008tw}, and write $\Psi$ as the Weyl fermion pair $(\chi, \eta^\dagger)$. Then the $\bar{\Psi^C} \Psi h_D^* h_D^*$ operator (and its conjugate) give rise to, 
\begin{equation} \mathcal{L}_\mathrm{split} = (1/2)(y/\Lambda)(\chi \chi h_D^* h_D^* + \eta \eta h_D h_D + h.c). \end{equation}
Consequently, once $h_D$ develops a VEV, we can write the mass terms for $\chi, \, \eta$ in the form,
\begin{equation} \mathcal{L}_\mathrm{mass} = \frac{1}{2} \left(\chi \quad \eta \right) \left(\begin{array}{cc} m_M & m_\chi \\ m_\chi & m_M \end{array} \right) \left(\begin{array}{c} \chi \\ \eta \end{array} \right) + h.c.,\end{equation} 
where $m_M = (y/\Lambda) \langle h_D \rangle^2$. We can now write down the Takashi diagonalization matrix $\Omega$ and the mass eigenstates $\chi_1, \chi_2$: 
\begin{equation} \Omega = \frac{1}{\sqrt{2}} \left(\begin{array}{cc} 1 & i \\ 1 & -i \end{array} \right), \quad \left(\begin{array}{c} \chi_1 \\ \chi_2 \end{array} \right) = \Omega^{-1} \left(\begin{array}{c} \chi \\ \eta \end{array} \right). \end{equation}
Then the mass matrix becomes,
\begin{align} \mathcal{L}_\mathrm{mass} & = \frac{1}{2} \left(\chi_1 \quad \chi_2 \right) \Omega^T \left(\begin{array}{cc} m_M & m_\chi \\ m_\chi & m_M \end{array} \right) \Omega \left(\begin{array}{c} \chi_1 \\ \chi_2 \end{array} \right) + h.c. \nonumber \\
& = \frac{1}{2} \left(\chi_1 \quad \chi_2 \right)  \left(\begin{array}{cc} m_\chi + m_M & 0 \\ 0 & m_\chi - m_M \end{array} \right) \left(\begin{array}{c} \chi_1 \\ \chi_2 \end{array} \right) + h.c.,\end{align}
as desired, and we see that the mass splitting $\delta = 2 m_M = 2 (y/\Lambda) \langle h_D \rangle^2$. The splitting term in $\mathcal{L}$ transforms to,
\begin{align} \mathcal{L}_\mathrm{split} & = (1/2)(y/\Lambda)((1/2) (\chi_1 + i \chi_2)^2 h_D^* h_D^* + (1/2) (\chi_1 - i \chi_2)^2 h_D h_D + h.c), \nonumber \\ & = (1/4) (y/\Lambda) (\chi_1 \chi_1 (h_D h_D + h_D^* h_D^*) - \chi_2 \chi_2 (h_D h_D + h_D^* h_D^*) \nonumber \\
& + 2 i \chi_1 \chi_2 (h_D h_D - h_D^* h_D^*)) + h.c. \end{align}
Working in unitarity gauge, we will write $h_D \rightarrow (v_D + \rho)/\sqrt{2}$, so $\delta = (y/\Lambda) v_D^2$; we then obtain,
\[ \mathcal{L}_\mathrm{split} \rightarrow (1/4) (y/\Lambda) (\chi_1 \chi_1 - \chi_2 \chi_2) (v_D^2 + \rho^2 + 2 v_D \rho) + h.c., \]
so the strength of the $\chi_i \chi_i \rho \rho$ interaction vertex is set by $y/\Lambda$ (the Yukawa interaction is suppressed by $v_D/\Lambda$). Furthermore, the mass of the gauge boson $\phi$ is given by $m_\phi^2/2 = g_D^2 \langle h_D \rangle^2$, so $m_\phi = g_D v_D$. Thus we have the relation,
\begin{equation} y/\Lambda = \delta / v_D^2 = \delta g_D^2 / m_\phi^2.\end{equation}

Computing the rate for $\chi_i \chi_i \rightarrow \rho \rho$ annihilation in the two-component formalism, we obtain,
\begin{equation} i \mathcal{M} = i \left(\frac{\delta g_D^2}{m_\phi^2}\right) \left(y_{1 \dot{\alpha}}^\dagger y_2^{\dagger \dot{\alpha}} + x_1^\alpha x_{2 \alpha} \right), \end{equation}
\begin{align} \frac{1}{4} \sum_{s_1,s_2} |\mathcal{M}|^2 & = \frac{1}{4}\left(\frac{\delta g_D^2}{m_\phi^2}\right)^2 \left(y_{1 \dot{\alpha}}^\dagger y_2^{\dagger \dot{\alpha}} + x_1^\alpha x_{2 \alpha} \right) \left(y_2^\beta y_{1 \beta} + x_{2 \dot{\beta}}^\dagger x_1^{\dagger \dot{\beta}} \right) \nonumber \\ 
& = \frac{1}{4}\left(\frac{\delta g_D^2}{m_\phi^2}\right)^2 \left(p_2 \cdot \bar{\sigma}^{\dot{\alpha} \beta} p_1 \cdot \sigma_{\beta \dot{\alpha}} +  p_1 \cdot \bar{\sigma}^{\dot{\beta} \alpha} p_2 \cdot \sigma_{\alpha \dot{\beta}} - m_\chi^2 \left(\delta_\beta^\alpha \delta_\alpha^\beta + \delta^{\dot{\alpha}}_{\dot{\beta}} \delta^{\dot{\beta}}_{\dot{\alpha}} \right) \right) \nonumber \\
& = \left(\frac{\delta g_D^2}{m_\phi^2}\right)^2 \left( p_1 \cdot p_2 - m_\chi^2 \right), \end{align}
exactly as previously. Since in the high-energy limit we can either say that half the particles are fermions and half antifermions, or that half are in the ``ground state'' and half in the ``excited state'' (really, the linear combinations of the Weyl fermions that will become the mass eigenstates at low energy), the $\Psi \Psi \rightarrow h_D h_D$ and $\psi_i \psi_i \rightarrow \eta \eta$ cross sections should indeed be equal to obtain a consistent overall annihilation rate.

Taking $\alpha_D = g_D^2 / 4 \pi$ as usual, we can write,
\begin{equation} \sigma = \frac{\pi \alpha_D^2}{4} \left(\frac{\delta }{m_\phi^2}\right)^2 \sqrt{1 - 4 m_\chi^2/s}. \end{equation}
In the low-energy limit $s \approx 4 m_\chi^2 (1 + v^2)$, and $v_\mathrm{rel} = 2 v$, so we obtain,
\begin{equation} \sigma v_\mathrm{rel} = \frac{1}{2} v^2 \left( \frac{\delta m_\chi }{m_\phi^2}\right)^2 \frac{\pi \alpha_D^2}{m_\chi^2}. \end{equation}

\bibliographystyle{JHEP}
\bibliography{benchmark}
\end{document}